\documentclass[fleqn,usenatbib]{mnras}

\usepackage{newtxtext,newtxmath}

\usepackage[T1]{fontenc}

\DeclareRobustCommand{\VAN}[3]{#2}
\let\VANthebibliography\thebibliography
\def\thebibliography{\DeclareRobustCommand{\VAN}[3]{##3}\VANthebibliography}

\usepackage{graphicx}	
\usepackage{amsmath}	

\usepackage{caption}
\usepackage{subcaption}

\usepackage{booktabs}
\usepackage{threeparttable}

\title[Discovery of a dynamically cold disc at $z=7.3$]{REBELS-25: Discovery of a dynamically cold disc galaxy at $z=7.31$}

\author[Lucie E. Rowland]{
Lucie E. Rowland$^{1}$\thanks{E-mail:  lrowland@strw.leidenuniv.nl},
Jacqueline Hodge$^{1}$,
Rychard Bouwens$^{1}$,
Pavel Mancera Piña$^{1}$,
Alexander Hygate$^{1}$,
\newauthor Hiddo Algera$^{2,3}$,
Manuel Aravena$^{4}$,
Rebecca Bowler$^{5}$,
Elisabete da Cunha$^{6,7}$,
Pratika Dayal$^{8}$,
\newauthor Andrea Ferrara$^{9}$,
Thomas Herard-Demanche$^{1}$,
Hanae Inami$^{2}$,
Ivana van Leeuwen$^{1}$,
Ilse de Looze$^{10}$,
\newauthor Pascal Oesch$^{11,12}$,
Andrea Pallottini$^{9}$,
Siân Phillips$^{13}$,
Matus Rybak$^{14,1,15}$,
Sander Schouws$^{1}$,
Renske Smit$^{13}$,
\newauthor Laura Sommovigo$^{16}$,
Mauro Stefanon$^{17,18}$,
Paul van der Werf$^{1}$
\\
$^{1}$ Leiden Observatory, Leiden University, P.O. Box 9513, 2300 RA Leiden,
The Netherlands\\
$^{2}$ Hiroshima Astrophysical Science Center, Hiroshima University, 1-3-1 Kagamiyama, Higashi-Hiroshima, Hiroshima 739-8526, Japan\\
$^{3}$ National Astronomical Observatory of Japan, 2-21-1, Osawa, Mitaka, Tokyo, Japan\\
$^{4}$ Núcleo de Astronomía, Facultad de Ingeniería y Ciencias, Universidad Diego Portales, Av. Ejército Libertador 441, Santiago, Chile \\
$^{5}$ Jodrell Bank Centre for Astrophysics, Department of Physics and Astronomy, School of Natural Sciences, The University of Manchester, Manchester M13 9PL, UK\\
$^{6}$ International Centre for Radio Astronomy Research, University of Western Australia, 35 Stirling Hwy, Crawley, WA 6009, Australia\\
$^{7}$ ARC Centre of Excellence for All Sky Astrophysics in 3 Dimensions (ASTRO 3D)\\
$^{8}$ Kapteyn Astronomical Institute, University of Groningen, 9700 AV Groningen, The Netherlands\\
$^{9}$  Scuola Normale Superiore, Piazza dei Cavalieri 7, 56126, Pisa, Italy\\
$^{10}$ Sterrenkundig Observatorium, Ghent University, Krijgslaan281-S9, B-9000 Ghent, Belgium\\
$^{11}$ Department of Astronomy, University of Geneva, Chemin Pegasi 51, 1290 Versoix, Switzerland\\
$^{12}$ Cosmic Dawn Center (DAWN), Copenhagen, Denmark\\
$^{13}$ Astrophysics Research Institute, Liverpool John Moores University, 146 Brownlow Hill, Liverpool L3 5RF, United Kingdom\\
$^{14}$ Faculty of Electrical Engineering, Mathematics and Computer Science, Delft University of Technology, Mekelweg 4, 2628 CD Delft, The Netherlands\\
$^{15}$ SRON - Netherlands Institute for Space Research, Niels Bohrweg 4, 2333 CA Leiden, The Netherlands\\
$^{16}$ Center for Computational Astrophysics, Flatiron Institute, 162 5th Avenue, New York, NY 10010, USA\\
$^{17}$ Departament d’Astronomia i Astrofísica, Universitat de València, C. Dr. Moliner 50, E-46100 Burjassot, València, Spain\\
$^{18}$ Unidad Asociada CSIC “Grupo de Astrofísica Extragaláctica y Cosmología” (Instituto de Física de Cantabria - Universitat de València)
}

\date{Accepted XXX. Received YYY; in original form ZZZ}

\pubyear{\the\year{}}

\begin{document}
\label{firstpage}
\pagerange{\pageref{firstpage}--\pageref{lastpage}}
\maketitle

\begin{abstract}
    We present high resolution ($\sim0.14$" = 710 pc) ALMA [CII] 158$\mu$m and dust continuum follow-up observations of REBELS-25, a [CII]-luminous ($L_{\mathrm{[CII]}}=(1.7\pm0.2)\times 10^9 \mathrm{L_{\odot}}$) galaxy at redshift $z=7.3065\pm0.0001$. These high resolution, high signal-to-noise observations allow us to study the sub-kpc morphology and kinematics of this massive ($M_* = 8^{+4}_{-2} \times 10^9 \mathrm{M_{\odot}}$) star-forming (SFR$_{\mathrm{UV+IR}} = 199^{+101}_{-63} \mathrm{M_{\odot}} \mathrm{yr}^{-1}$) galaxy in the Epoch of Reionisation. By modelling the kinematics with \texttt{$^{\mathrm{3D}}$BAROLO}, we find it has a low velocity dispersion ($\bar{\sigma} = 33 \pm 9$ km s$^{-1}$) and a high ratio of ordered-to-random motion ($V_{\mathrm{rot, ~max}}/\bar{\sigma} = 11 ^{+8}_{-4}$), indicating that REBELS-25 is a dynamically cold disc. Additionally, we find that the [CII] distribution is well fit by a near-exponential disc model, with a Sérsic index, $n$, of $1.3 \pm 0.2$, and we see tentative evidence of more complex non-axisymmetric structures suggestive of a bar in the [CII] and dust continuum emission. By comparing to other high spatial resolution cold gas kinematic studies, we find that dynamically cold discs seem to be more common in the high redshift Universe than expected based on prevailing galaxy formation theories, which typically predict more turbulent and dispersion-dominated galaxies in the early Universe as an outcome of merger activity, gas accretion and more intense feedback. This higher degree of rotational support seems instead to be consistent with recent cosmological simulations that have highlighted the contrast between cold and warm ionised gas tracers, particularly for massive galaxies. We therefore show that dynamically settled disc galaxies can form as early as 700 Myr after the Big Bang.

\end{abstract}




\section{Introduction}

\label{sec:intro}

In the current picture of galaxy formation, galaxies in the early Universe should be more dominated by turbulent motion than their lower redshift counterparts, as a result of an increase in merger activity and more violent gas accretion (\citealt{conselice_structures_2008,dekel_cold_2009}), as well as more intense active galactic nuclei (AGN) and stellar feedback (e.g., \citealt{hayward_how_2017,nelson_first_2019}). With the advent of the Hubble Space Telescope (HST), high angular resolution rest-frame UV images of redshift $z>1$ galaxies indeed revealed more clumpy and irregular morphologies (e.g., \citealt{conselice_evidence_2003,papovich_assembly_2005,elmegreen_stellar_2005,conselice_structures_2008}), in comparison to the rotationally supported disc galaxies that make up 50-80\% of the $z\sim$ 0-1 Universe (e.g., \citealt{kassin_epoch_2012,wisnioski_kmos3d_2015,swinbank_angular_2017}). In addition, multiple integral field (IFU) surveys of warm ionised gas kinematics have found that the turbulence (as measured by the velocity dispersion, $\sigma$, of the gas) within galaxies increases from $z=0$ to $z \sim 3$, and that the degree of rotational support (rotational velocity over velocity dispersion, $V_{\mathrm{rot}}/\sigma$) decreases over the same redshift range (e.g., \citealt{wisnioski_kmos3d_2015,di_teodoro_flat_2016, johnson_kmos_2018,wisnioski_kmos3d_2019,ubler_evolution_2019}).

However, these findings have recently been challenged on multiple fronts. JWST, with its unprecedented high-angular resolution and sensitivity in the near- and mid-infrared, has now revealed more massive ($M_*\gtrsim 10^{9}-10^{10} \mathrm{M_{\odot}}$) disc-like galaxies at $z>6$ (\citealt{adams_discovery_2022,akins_two_2023,labbe_population_2023,rodighiero_jwst_2023,casey_cosmos-web_2023, atek_jwst_2023}) than predicted from HST observations, hinting that significant disc formation may have occurred much earlier than expected. Similarly, both JWST and the Atacama Large Millimetre/submillimetre Array (ALMA) have observed an increasing number of dynamically cold ($V_{\mathrm{rot}}/\sigma > 2 $) disc galaxies at $z>3$ (\citealt{smit_rotation_2018}; \citealt{sharda_testing_2019}; \citealt{neeleman_cold_2020,rizzo_dynamically_2020,neeleman_kinematics_2021,lelli_massive_2021,rizzo_dynamical_2021,fraternali_fast_2021,tsukui_spiral_2021,roman-oliveira_regular_2023,posses_structure_2023,pope_alma_2023, neeleman_alma_2023, de_graaff_ionised_2023,parlanti_ga-nifs_2023,ubler_ga-nifs_2024, xu_dynamics_2024}). Cosmological simulations often struggle to reproduce these findings, with simulated rotationally supported gas discs typically only forming at $z\lesssim 2$ (\citealt{simons_distinguishing_2019,pillepich_first_2019}).

These discrepancies and recent discoveries have made the early build-up of galaxies a forefront challenge in extragalactic astronomy today. To address the tension between observations and theory, some studies over the past few years have suggested that the `disciness' of galaxies is more dependent on the mass of the halo, rather than on the redshift (\citealt{dekel_mass_2020,gurvich_rapid_2022}; \citealt{tiley_kmos_2021}; \citealt{kohandel_dynamically_2023}), with gas discs only surviving in $M_*\gtrsim 10^9 \mathrm{ M_{\odot}}$ haloes. However, discrepancies remain between the previous findings of an increase in turbulence with redshift, in comparison to the growing number of discoveries of dynamically cold discs at $z>3$. Likely explanations could include the spatial resolution, signal-to-noise ratio (SNR), sample selection, and kinematic tracer used for observations (e.g. \citealt{rizzo_dynamical_2022,kohandel_dynamically_2023}). It has therefore become clear that deep, high spatial resolution, multi-wavelength observations of a range of galaxy types are critical to fully investigate the morphological and kinematic evolution of galaxies, and thus inform how galaxies build up their mass over cosmic time. ALMA has been fundamental in such investigations, in particular thanks to detections of the [CII] (157.7$\mu$m) far-infrared (FIR) emission line (e.g. \citealt{stacey_optical_1991, malhotra_infrared_1997}, and see review of \citealt{hodge_high-redshift_2020}). This line is one of the brightest interstellar medium (ISM) cooling lines, tracing the cold ($T \sim 100$K) neutral, molecular and ionised ISM (\citealt{shibai_cii_1996,stacey_158_2010,pineda_herschel_2013,vallini_cii-sfr_2015}).
 
The Reionization Era Bright Emission Line Survey (REBELS, \citealt{bouwens_reionization_2022}) is an ALMA large program (LP) that has carried out spectral scans for the [CII] line in ALMA's Band 6 for $\sim$ 40 UV-bright star-forming galaxies at $z\sim 6-8$. As well as multiple, significant [CII] detections, this LP has also enabled simultaneous dust continuum detections of these galaxies (\citealt{inami_alma_2022}). This sample therefore presents an opportunity to study mass assembly during the Universe's last major phase transition; the Epoch of Reionization (EoR). This program has highlighted ALMA's capabilities as a `redshift machine' (see also \citealt{smit_rotation_2018, schouws_significant_2022}), as well as enabling studies of, for example, dust-obscured star formation (\citealt{fudamoto_normal_2021}), specific star formation rates (sSFR, \citealt{topping_alma_2022}), Ly$\alpha$ emission (\citealt{endsley_alma_2022}), and dust and ISM properties (\citealt{sommovigo_rebels_2022, dayal_alma_2022, ferrara_alma_2022}) at $z>6$. Some of these REBELS galaxies show evidence of velocity gradients in the LP data (Schouws et al. in prep), including REBELS-25, which is the most infrared luminous and most promising rotating disc candidate from the REBELS LP (\citealt{hygate_alma_2023}). However, at the relatively low ($\gtrsim$ 1") resolution of the LP observations,  morphological and kinematic classification of galaxies remains challenging (e.g., \citealt{goncalves_kinematics_2010,simons_distinguishing_2019,rizzo_dynamical_2021}). Indeed, in \cite{hygate_alma_2023}, the angular resolution of the LP observations of REBELS-25 ($\sim 1.3$" $= 6.7$ kpc)  was found to be insufficient to distinguish between a rotating disc and merger scenario. In addition, analysis of HST rest-frame UV observations by \cite{stefanon_brightest_2019} revealed a clumpy UV morphology for this galaxy, which could be indicative of a merger. For these reasons, follow-up high resolution [CII] and dust continuum observations have been obtained for REBELS-25 (ID 2021.1.01603.S, PI J. Hodge) and planned/executed for a subsample of other [CII]-luminous REBELS galaxies (ID 2022.1.01131.S, PI R. Smit).

In this work, we present the morphological and kinematic analysis at sub-kpc resolution for REBELS-25. Analysis of the remaining high resolution REBELS subsample will be the focus of subsequent works (Phillips et al in prep). In Section \ref{sec:observations}, we discuss the observations and the data reduction used in this high resolution follow-up analysis of REBELS-25. In Section \ref{sec:morphology}, we present the morphological analysis of REBELS-25, and then in Section \ref{sec:kinematic modelling set up} we present the methodology used to investigate its kinematics, with the best-fit models and tests used to distinguish between the merger and rotating disc scenario discussed in Section \ref{sec:kinematic analysis}. In Section \ref{sec:discussions}, we discuss and interpret our findings and make comparisons to other similar studies, and finally in Section \ref{sec:conclusions} we summarise our main results and their implications. Throughout the paper, we adopt a standard $\Lambda$CDM cosmology with Hubble constant $H_0=70$ km s$^{-1}$ Mpc$^{-1}$, matter density $\Omega_m=0.3$ and vacuum energy $\Omega_{\Lambda}=0.7$. At the redshift of REBELS-25, this corresponds to a luminosity distance of 72519 Mpc. 

\begin{table}
\caption[]{Table showing the properties of REBELS-25 from previous studies.}
\normalsize
\centering
\begin{tabular}{lll}
\centering
        
        Property & Value & Reference \\
	\hline\hline
	$z_{\mathrm{[CII]}}$  &  7.3065 $\pm$ 0.0001 &  1,2  \\
	
	$M_*$ ($M_{\odot}$) & $8^{+4}_{-2} \times 10^9$ & 2, 3 \\
	
        $M_{H_2,[CII]}$ ($M_{\odot}$) & $5.1^{+5.1}_{-2.6} \times 10^{10}$ & 4\\
        
        SFR$_{\mathrm{[CII]}}$ ($M_{\odot}\mathrm{yr}^{-1}$) & 246 $\pm$ 35 & 4 \\
        
        SFR$_{IR}$ ($M_{\odot}\mathrm{yr}^{-1}$) & $185 ^{+101}_{-63}$ & 5\\
        
        $SFR_{UV}$($M_{\odot}\mathrm{yr}^{-1}$) & $14 \pm 3$ & 3 \\
        
        $L_{[CII]}$ ($L_{\odot}$) & 1.7 $\pm$ 0.2 $\times 10^9$ & 4 \\
        
        $L_{[IR]}$ ($L_{\odot}$) & $7.1 ^{+3.6}_{-1.5} \times 10^{11}$ & 6 \\
        
        $S_{158\mu m}$ ($\mu$Jy) & $260 \pm 22$ & 5  \\
        
        FWHM$_{[CII]}$ (km s$^{-1}$) & $316 \pm 15$ & 4\\
        
        Conversion (kpc/") & 5.09552\\
        \hline\hline

\end{tabular}
\begin{tablenotes}
\item References: [1] Schouws et al. in prep, [2] \cite{bouwens_reionization_2022}, [3] Stefanon et al. (in prep), [4] \cite{hygate_alma_2023}, [5] \cite{inami_alma_2022}, [6] \cite{algera_cold_2023}.
\end{tablenotes}
\label{tab:R25 properties}

\end{table}

\section{Observations}
\label{sec:observations}

REBELS-25, also known as UVISTA-Y-003 or UVISTA-Y3 (\citealt{stefanon_brightest_2019,schouws_significant_2022}), has J2000 RA, Dec coordinates of $10^{\mathrm{h}} 00^{\mathrm{m}} 32.32^{\mathrm{s}}$, $+01^{\circ}44'31.3$" from rest-frame UV imaging (\citealt{stefanon_brightest_2019}) and a [CII] spectroscopic redshift of $z=7.3065\pm0.0001$ (\citealt{bouwens_reionization_2022}). Analysis of the REBELS LP data of this galaxy has been detailed in \cite{hygate_alma_2023}. Here, we present the sub-kpc resolution observations used to further investigate the nature of this source.

\subsection{ALMA Data reduction and imaging}
\label{sec:reduction}

The high resolution Band 6 observations of REBELS-25 were obtained in ALMA Cycle 8 with ID 2021.1.01603.S (PI J. Hodge). These observations were carried out in November 2021 in the C-6 configuration, under good weather conditions (typical precipitable water vapour of 0.4 mm). The baseline lengths spanned between 41.4 m and 3.6 km, resulting in a maximum recoverable scale (MRS) of $\sim 1.7$" and an angular resolution of $\sim 0.1$". In total, 256.67 minutes were spent on source. The calibration and flagging were done using the standard pipeline running on \texttt{CASA} (Common Astronomy Software Applications for Radio Astronomy, \citealt{the_casa_team_casa_2022}) version 6.2.1.7, and we use the same pipeline version for all subsequent imaging. Inspection of the pipeline-calibrated data tables revealed data of high quality, and the $uv$-data were therefore used without further modification to the calibration scheme or flagging. This Cycle 8 data set contains a spectral window covering the [CII] line with a bandwidth of 1875 MHz, and three other spectral windows for detecting the FIR dust continuum surrounding the line ($\sim 150 \mu$m), each with a bandwidth of 2000 MHz.

To obtain a [CII] line-emission cube, we first created a continuum-free measurement set using the task \texttt{uvcontsub} with a zeroth-order fit to line-free channels, excluding a region that is two times the full width at half maximum (FWHM) of the [CII] line. We created a dirty line-emission cube using \texttt{tclean} in the `cube’ mode to find the root mean square (RMS) noise level per velocity channel and then produced a final line-emission cube cleaned down to two times this RMS noise. This cleaning was done iteratively using \texttt{CASA}'s automask mode with the sub-parameters initially determined by the recommendations of the ALMA automasking guide\footnote{\url{https://casaguides.nrao.edu/index.php/Automasking_Guide}} for long baselines. On inspection of the mask created for the cleaning and the residuals produced, we lowered the `noisethreshold' parameter to 4 and the `sidelobethreshold' to 1.5 (both in units of the RMS) to ensure cleaning of source emission in channels not masked in the preliminary clean. The rest of the sub-parameters are left at the recommended values (`lownoisethreshold' = 1.5, `negativethreshold' = 7, `minbeamfrac' = 0.3). The native channel width of the data is $\sim$ 5.1 km s$^{-1}$. In the final line-cube, we bin the channels by a factor of 3 to a channel width of $\sim 15.4$ km s$^{-1}$, as a compromise between sensitivity and velocity resolution.

All imaging was carried out with a multi-scale \texttt{CLEAN}; a modification of the classical \texttt{CLEAN} algorithm that assumes sources in the sky are extended structures of different scales. We use a small-scale bias of 0.9 and deconvolution scales of 0, 1 $\times$ the beam FWHM, and 3 $\times$ the beam FWHM. We find that the use of this multi-scale \texttt{CLEAN}, in comparison to the default `hogbom' deconvolver, is more successful at masking both compact and extended emission during the cleaning process.

A $\sim150 \mu$m continuum map was similarly produced from the channels not containing line or atmospheric emission in the `mfs' mode. The cleaning was also done iteratively with automasking and cleaned down to two times the RMS noise of the dirty continuum image. We will refer to this continuum map as the dust continuum hereafter.

With the cleaning process described above, we produced both natural-weighted and Briggs-weighted line cubes and continuum maps. The Briggs weighted images were obtained with a robust parameter of 0.5, which is a compromise between sensitivity and spatial resolution. For the [CII] imaging, the natural-weighted beam size is 0.14"$\times$0.13" (710$\times$660 pc), and the Briggs-weighted beam size is 0.12"$\times$"0.10 (610$\times$510 pc). For the dust continuum, the natural-weighted beam size is 0.14"$\times$0.13" (710$\times$660 pc), and the Briggs-weighted beam size is 0.11"$\times$0.09" (560$\times$460 pc). Although the Briggs-weighted imaging achieves a higher spatial resolution, the increased sensitivity of the natural weighted data allows for a higher signal-to-noise, particularly in the outer regions of the galaxy, and results in more resolution elements across the source, whilst still providing sub-kpc resolution. For the subsequent analysis, we therefore focus on the naturally weighted imaging. For the [CII] line-cube, the final RMS per 15 km s$^{-1}$ channel is $\sim110 \mu$Jy beam$^{-1}$. For the dust continuum map, the final RMS is $\sigma\sim$5.7 $\mu$Jy beam$^{-1}$. We find consistent [CII] and dust continuum flux densities, within the uncertainties, when comparing the high resolution data with the low resolution REBELS LP data.

Whilst it would be possible to increase the sensitivity of these data by concatenating the high resolution data set with the low resolution LP data, the combination of very different array configurations means we would be impacted by the "JvM effect" (see \citealt{jorsater_high_1995, czekala_molecules_2021,posses_alma-cristal_2024}). Whilst it is possible to correct for this effect (\citealt{czekala_molecules_2021}), this correction means that the residuals have been rescaled, and so the RMS is no longer a true representation of the sensitivity of the observations. The findings of \cite{casassus_variable_2022}, in particular, caution against the use of this JvM correction. Since the focus of this paper is a morphological and kinematic analysis, with tools dependent on some SNR masking, these rescaled residuals would impact all subsequent analysis. For these reasons, we choose not to combine the low and high resolution imaging, although we discuss tests using a JvM-corrected, concatenated [CII] line cube (as well as line cubes with different weighting schemes and different channel binnings) in Appendix A. Overall, we find that the main results and conclusions of this paper are robust against the data set and reduction process used.

\subsection{Rest-frame UV imaging}

In order to compare the [CII] and dust continuum morphology with the rest-frame UV emission, we also make use of the HST Wide Field Camera 3 (WFC3) F160W imaging from the COSMOS-DASH survey (PID 13868, PI D. Kocevski). Details of the HST imaging extracted from the COSMOS-DASH mosaic (\citealt{mowla_cosmos-dash_2019}) and the astrometry correction are given in \cite{hygate_alma_2023}.

\section{Morphology}
\label{sec:morphology}

\begin{figure*}
    \centering
    \includegraphics[width=0.98\textwidth]{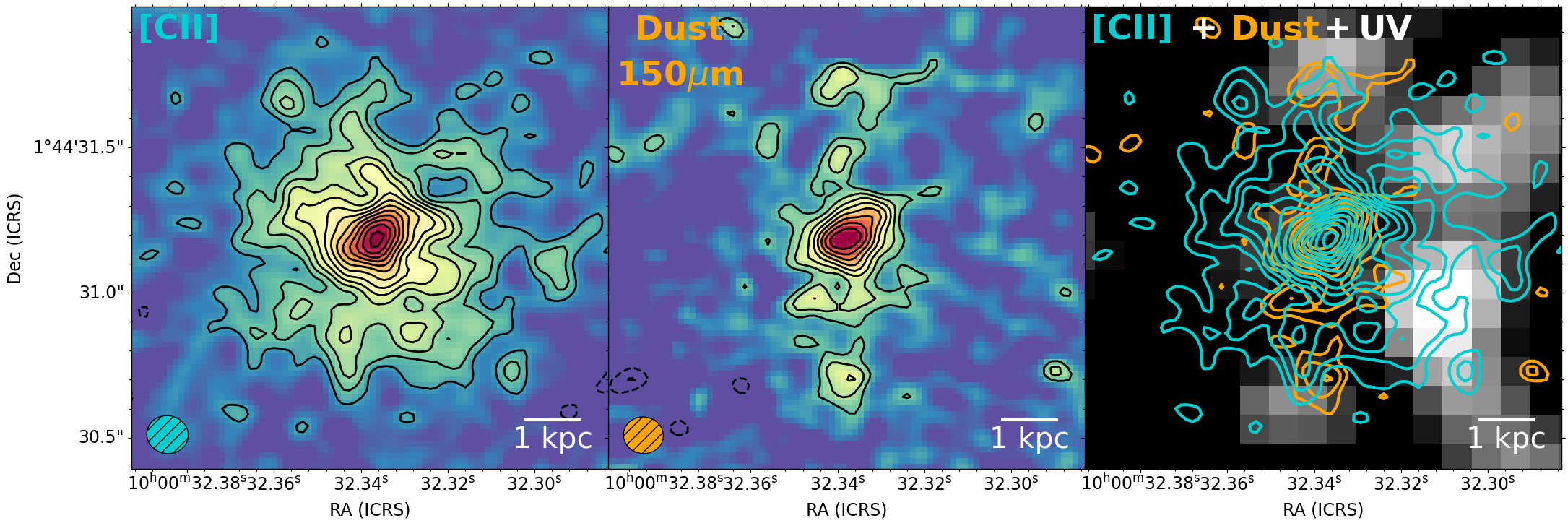}
    \caption {\textit{Left}: [CII] moment-0 map from the naturally weighted data cube, with contours showing 2, 3, ...14$\sigma_{\mathrm{RMS}}$ emission (where $\sigma_{\mathrm{RMS}}=11$ mJy beam$^{-1}$ km s$^{-1}$). The size of the beam is indicated by the turquoise ellipse in the bottom left corner. \textit{Centre}: Naturally weighted dust continuum map, with orange contours showing 2, 3, ...10$\sigma_{\mathrm{RMS}}$ emission (where $\sigma_{\mathrm{RMS}}=5.5 ~\mu$Jy beam$^{-1}$). The size of the beam is indicated by the orange ellipse in the bottom left corner. \textit{Right}: HST WFC3 F160W image from the COSMOS-DASH mosaic (\citealt{mowla_cosmos-dash_2019}) with the [CII] emission and dust continuum shown by the turquoise and orange contours, respectively.}
    \label{fig:presenting data}
\end{figure*}

In Figure \ref{fig:presenting data}, we present the [CII], dust and rest-frame UV morphology of REBELS-25 at sub-kpc resolution. The [CII] emission reveals a clear extended disc, whereas the rest-frame UV morphology can be separated into clumps that are offset from the [CII] and dust continuum. Both the [CII] and dust maps show an inner, bright region that is misaligned from the extended gas disc. In the following sections, we detail the fitting tools used to investigate these morphological features.

\subsection{[CII] morphology}
\label{sec: cii morphology}

To investigate the morphology of the [CII] emission in REBELS-25, we use the CASA task \texttt{immoments} to create a moment-0 map by integrating a spectral region covering 2 $\times$ FWHM of the [CII] line emission for the naturally weighted line cube (Figure \ref{fig:presenting data}). From this moment-0 map, we find a peak SNR of 14.3. We then fit a 2D Sérsic profile to the moment-0 map using the Astropy \texttt{Sersic2D} modelling class and the task \texttt{TRFLSQFitter} with the models convolved with the beam of the observations using \texttt{PetroFit} (\citealt{2022AJ....163..202G}). From this fitting, we find a position angle of $224\pm28^{\circ}$, an effective radius, $r_{\mathrm{e}}$, of 0.42" $\pm$ 0.06" ($2.1\pm0.2$ kpc), and a Sérsic index, $n$, of $1.3 \pm 0.2$, indicating that the [CII] emission is well-fit by a near-exponential disc. The best-fit ellipticity is 0.13$\pm$0.09, which corresponds to an inclination, $i$, of $30^{+10}_{-15}$$^{\circ}$, assuming the disc is razor-thin. However, previous studies have found that high-$z$ galaxies are likely to have thicker discs due to increased turbulence. The assumption of a thin disc would therefore result in an underestimation of the inclination, which subsequently impacts the inclination-corrected velocities.

We therefore use the \texttt{CANNUBI}\footnote{\url{https://www.filippofraternali.com/cannubi}} package to also fit for the thickness and inclination of the disc. As introduced in \cite{pina_robust_2020,pina_no_2022,fraternali_fast_2021} and \cite{roman-oliveira_regular_2023}, \texttt{CANNUBI} fits the inclination ($i$), the disc thickness ($Z_0$), the morphological centre ($x_{0,\mathrm{morph}},y_{0,\mathrm{morph}}$), and morphological position angle ($PA_\mathrm{morph}$) of galactic discs. A more detailed description of \texttt{CANNUBI} is given in \cite{roman-oliveira_regular_2023}, but in summary, we use \texttt{CANNUBI} to fit the observed [CII] flux map with resolution-matched maps of 3D tilted-ring model disc galaxies. This means that \texttt{CANNUBI} models the galaxy morphology without a prior assumed parametric description of the surface brightness distribution.  We set a lower limit of 15$^\circ$ to the fitting for the inclination, which is a reasonable lower limit based on the inclination derived above from the ellipticity fitted using \texttt{Sersic2D}, assuming a razor-thin disc. The best-fit parameters to the disc geometry are returned via a Markov Chain Monte Carlo (MCMC) routine with 30 walkers and run until convergence, as estimated according to the auto-correlation times of the parameters (\citealt{goodman_ensemble_2010}). The posterior distributions from the \texttt{CANNUBI} fitting to REBELS-25 are shown in Figure \ref{fig:cannubi}, and we list the median values and their uncertainties (from the 16th and 84th percentiles) in Table \ref{tab:R25 parameters}.

We find that, overall, the [CII] morphological parameters are well constrained via the \texttt{CANNUBI} fitting procedure. One exception is the fitting for $Z_0$, from which we can only infer that the best-fit models have a thickness lower than the $\sim 0.14"$ (710 pc) resolution of our observations.  From tests carried out on mock galaxies in \cite{roman-oliveira_regular_2023}, we anticipate that the poorly-fitted disc thickness will not significantly impact the derived inclination, since this effect is found to be more important for thicker discs. Having confirmed the thin disk assumption, we therefore set $Z_0 = 0$ for the subsequent modelling. The posterior for the inclination also appears to have a tail extending up to the 15$^\circ$ lower limit, however this is likely due to the \texttt{CANNUBI} models appearing very similar for low inclination galaxies (\citealt{pina_no_2022}). We note that if the inclination is in fact lower, this would only increase the derived intrinsic rotational velocities and further strengthen our findings, as discussed in Section \ref{sec:kinematic analysis}. The $PA_{\mathrm{morph}}$ posterior also shows a secondary, lower peak at $\sim 160 ^\circ$, which is potentially caused by a bright, central, misaligned component which we discuss in Section \ref{sec:potential features}.

The morphological centre and $PA_{\mathrm{morph}}$ returned via the Sérsic and \texttt{CANNUBI} models are consistent within errors. \texttt{CANNUBI} also returns the surface density within each ring, or resolution element. We find that this surface brightness profile is best fit with a 1D Sérsic profile with $n = 1.13 \pm 0.06$, which is also consistent with the 2D Sérsic fit. The inclination derived from \texttt{CANNUBI}, $i=25\pm6^{\circ}$, is lower but within error of the $i=56\pm29^{\circ}$ value determined from the ratio of the major and minor axes of the low resolution LP data in \cite{hygate_alma_2023}, where the [CII] emission is only marginally resolved and therefore the axis ratio is strongly dependent on the beam. As mentioned above, the inclination has a significant impact on the derived kinematic properties, highlighting the necessity of high spatial resolution observations for both morphological and kinematic analyses. 

In both the \texttt{CANNUBI} and \texttt{Sersic2D} fitting, we see some residual ($\sim 6\sigma_{\mathrm{RMS}}$ in the \texttt{CANNUBI} fitting) emission at the centre of the [CII] map. This feature is discussed in more detail below, in Section \ref{sec:potential features}. Despite these residuals, the \texttt{CANNUBI} model produces a better reduced chi squared value ($\chi^2 \sim 0.9$ for \texttt{CANNUBI}) than the Sérsic model ($\chi^2 \sim 0.18$ for \texttt{Sérsic2D}). Additionally, \texttt{CANNUBI} uses the same software as will be used for our kinematic analysis in Section \ref{sec:kinematic modelling set up}. For these reasons, we adopt the morphological parameters for the [CII] emission returned by \texttt{CANNUBI} as our fiducial values, and list these in Table \ref{tab:R25 parameters}.

\begin{figure}
    \centering
    \includegraphics[width=0.48\textwidth]{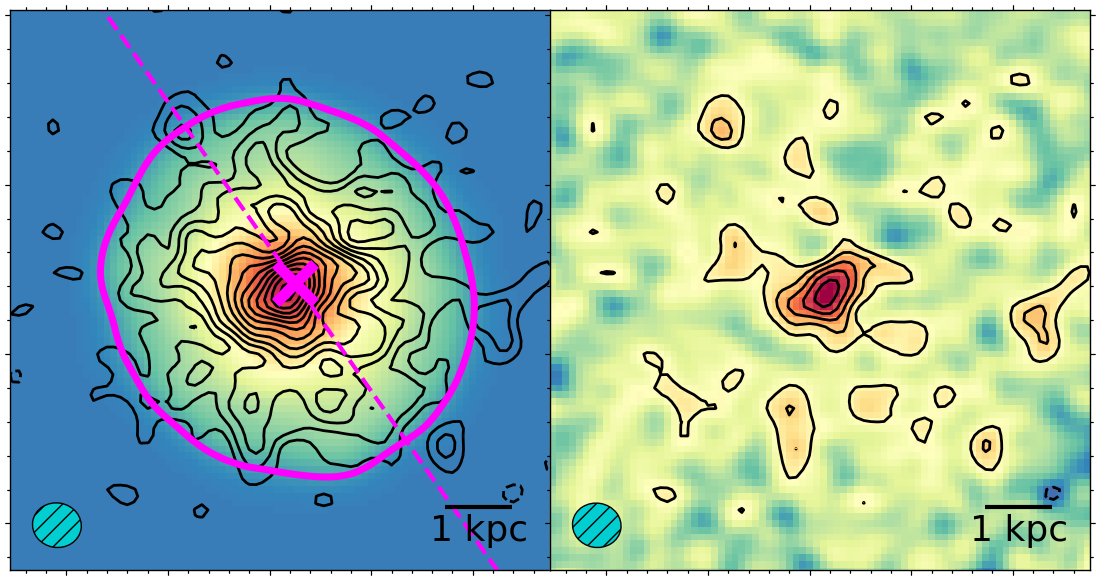}
    \caption {\textit{Left}: We show the best-fit \texttt{CANNUBI} model, with the observed [CII] emission from Figure \ref{fig:presenting data} overplotted in black contours. The dashed magenta line, magenta cross and magenta ellipse mark the morphological position angle, the centre and the radial extent of the best-fit model, respectively. \textit{Right}: The residuals from the best-fit \texttt{CANNUBI} model, with 2, 3, ..., 6$\sigma_{\mathrm{RMS}}$ contours shown by the solid black lines, and -2, -3$\sigma_{\mathrm{RMS}}$ contours show by the dashed black lines, where $\sigma_{\mathrm{RMS}}$ is the local RMS in the [CII] moment-0 map.}
    \label{fig:cannubi model}
\end{figure}

\begin{figure}
    \centering
    \includegraphics[width=0.48\textwidth]{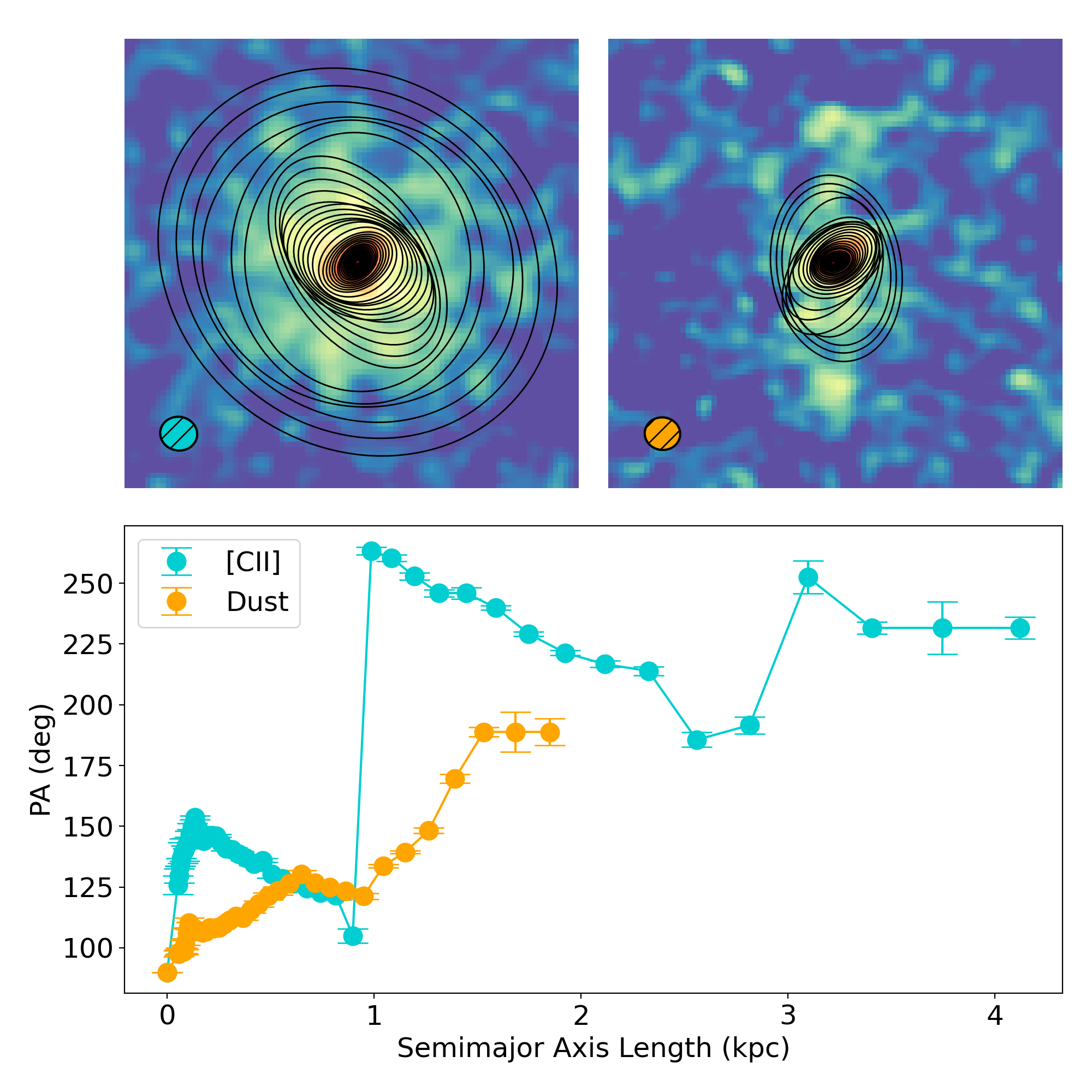}
    \caption {\textit{Top left}: The elliptical isophote fit to the [CII] emission is shown by the black contours. \textit{Top right}: The elliptical isophote fit to the dust continuum map. \textit{Bottom}: The position angle of the fitted ellipses as a function of radius for the [CII] (turquoise) and dust (orange) emission.}
    \label{fig:elliptical fit}
\end{figure}

\subsection{Dust continuum morphology}

In Figure \ref{fig:presenting data}, we also show the dust continuum map of REBELS-25, where we obtain a peak SNR of 10.1. As with the [CII] morphology, we also fit a 2D Sérsic model to the dust continuum and find $r_e=0.50$" $\pm 0.08$" ($2.5\pm0.7$ kpc) and $n=2.2\pm0.4$, indicating emission that is more centrally peaked than would be expected for an exponential disc.  We find $\sim 3\sigma$ residuals in clumps to the North and South of the centre from the \texttt{Sersic2D} fitting. These clumps are around 0.5" ($\sim 2.5$ kpc) from the fitted centre, and each contain around 10\% of the flux of the central component.

\subsection{Comparing the [CII], dust and rest-frame UV morphology}
\label{sec:potential features}

From the lower resolution REBELS LP data, a potential offset of $0.17" \pm 0.04"$ (around $0.9 \pm 0.2$ kpc) between the [CII] emission and the dust continuum is identified in \cite{hygate_alma_2023}. With the follow-up, sub-kpc resolution observations analysed here, however, we find that the centres of the dust and [CII] emission are in good agreement, again highlighting the necessity of observations with sufficient angular resolution. 

The [CII] morphology is best described by an exponential disc, whereas the dust morphology is not as well fit by a single component Sérsic model, likely due to the two faint clumps to the North and South of the peak of emission. These clumps drive the $PA_{\mathrm{morph}}$ of the [CII] and dust to differ by $\sim 60^{\circ}$ at the outer extent of emission. 

We also find that the effective radii of the [CII] and dust emission are comparable, with the $r_e$ of the dust even slightly larger (although they are consistent within the uncertainties). This is in contrast with other studies of high redshift star forming galaxies where the dust emission is found to be around  $2\times$ more compact than the [CII] emission (e.g., \citealt{fudamoto_alma_2022, fujimoto_alpine-alma_2020}). However, we note that the Sérsic model is a poor fit to the dust emission, and this large $r_e$ is likely driven by the clumpy structures to the North and South, and we therefore caution against comparing the effective radii of these Sérsic fits.

Notably, both the [CII] and dust show a central region in the inner $\sim$ 1 kpc (around 1.4-2 beams wide) that appears misaligned with the $PA_{\mathrm{morph}}$ of the extended gas disc. We see also see this misaligned, central region in the residuals around the fitted centre of the galaxy in both the [CII] and dust continuum 2D fits, as well as a tail in the posterior distribution for PA$_{\mathrm{morph}}$ with \texttt{CANNUBI}. To illustrate this, we fit elliptical isophotes to the [CII] moment-0 and dust map using the `isophote.Ellipse.fit\_image' in \texttt{PHOTUTILS} from the Astropy package (\citealt{bradley_astropyphotutils_2020}) with a sigma clip of 2$\sigma_{\mathrm{RMS}}$. For the [CII] fit, we follow the methodology outlined in \cite{amvrosiadis_onset_2024}, that is, we first leave all parameters free, and then for the final iteration we fix the centre of all ellipses to the median value. However, the fit for the dust fails to converge when the centre is fixed, and we therefore leave all parameters free for the dust fitting. The resulting fits are shown in Figure \ref{fig:elliptical fit}.

Within the central 1 kpc, the ellipses fitted to the [CII] and dust have an average PA of $\sim 120^{\circ}$. At roughly the same radii ($\sim 1$ kpc), the PA for the [CII] suddenly increases to an average PA of $\sim 220^{\circ}$ at $\gtrsim 1$ kpc (consistent with $PA_{\mathrm{morph}}$ from \texttt{CANNUBI} and \texttt{Sersic2D}), and for the dust it changes more gradually to $\sim190^{\circ}$ (consistent with $PA_{\mathrm{morph}}$ from \texttt{Sersic2D} for the dust). In addition, the ellipticity in the central 1 kpc in both the [CII] and dust is found to be $\sim 0.3$, whereas for the extended [CII] disc this is found to be $\sim 0.1$ (near-circular). 

Similar changes in both the ellipticity and position angle from elliptical isophote fits have been used to identify and analyse bars within nearby galaxies (e.g., \citealt{abraham_evolution_1999, laine_nested_2002, jogee_bar_2004, menendez-delmestre_near-infrared_2007, consolandi_automated_2016}), and now even for some higher redshift galaxies with JWST (e.g., \citealt{huang_j0107a_2023} at $z=2.467$, \citealt{amvrosiadis_onset_2024} at $z=3.762$, \citealt{leconte_jwst_2024} at $1\leq z\leq3$). However, we note that this tentative bar-like central component in REBELS-25 is barely resolved with these observations.

The rest-frame UV morphology differs significantly from the [CII] and dust. As previously identified in \cite{hygate_alma_2023} and \cite{schouws_significant_2022}, the HST imaging reveals 3-4 clumps that are $\sim0.6$" offset from the centre of both the [CII] and dust continuum emission. We see in Figure \ref{fig:presenting data} that these UV clumps lie close to and outside the outer edge of the [CII] and dust emission. As discussed in \cite{hygate_alma_2023}, these UV clumps could have a number of explanations, including differential dust obscuration, regions of intense star formation or potential merger activity. We further discuss these possible interpretations in Section \ref{sec:discussion1}.

\section{Kinematics}

We next focus on the [CII] kinematics of REBELS-25. In Section \ref{sec:kinematic modelling set up}, we first provide an overview of the techniques used to investigate the kinematic properties. We then present the results of our kinematic modelling in Section \ref{sec:kinematic analysis}, along with some investigation of the potential mechanisms that could explain the observed properties.

\subsection{Methods}
\label{sec:kinematic modelling set up}

For the kinematic modelling of REBELS-25, we use \texttt{$^{\mathrm{3D}}$BAROLO} (\citealt{di_teodoro_3d-barolo_2015}, v1.7). \texttt{$^{\mathrm{3D}}$BAROLO} is well-tested at a range of redshifts and from low to high resolution observations (e.g. at low resolution: \citealt{di_teodoro_3d-barolo_2015,pina_off_2019, gray_catching_2023}, at mid resolution: \citealt{bacchini_evidence_2020,pina_impact_2022, roman-oliveira_regular_2023}, and at high resolution: \citealt{iorio_little_2016,di_teodoro_radial_2021}). Its algorithm fits 3D tilted ring models to emission line data cubes on a channel-by-channel basis, and uses Monte Carlo methods to return the best-fit kinematic and/or morphological parameters. The tilted ring models are convolved to the same resolution as the input observations, and it therefore accounts for the effects of beam smearing. 

Before the modelling, we extract a 3.5"$\times$3.5" sub-cube centred on the [CII] emission that covers $\pm 4\times$ the FWHM of the [CII] emission in the spectral dimension from our full data cube. We then use the \texttt{SEARCH} algorithm to carry out a source scan on the inputted sub cube, with a primary SNR cut of $3\sigma_{\mathrm{RMS}}$. To grow a mask around the detected source, a range of \texttt{GROWTHCUT} parameters were tested. Based on these preliminary tests, we find a threshold of 2.2$\times 10^{-4}$ Jy beam$^{-1}$, which is is equivalent to 2$\sigma_{\mathrm{RMS}}$, is sufficient to produce a mask that encloses all of the [CII] emission without introducing too much noise. We set the radial separation of the rings to 0.11" ($\sim$80\% of the beam FWHM), resulting in five rings within the mask produced by \texttt{$^{\mathrm{3D}}$BAROLO}. This maximises the number of rings whilst ensuring each ring is nearly comparable to an independent resolution element. For all fits with \texttt{$^{\mathrm{3D}}$BAROLO}, we use an azimuthal normalisation of the surface density, we fit the kinematics using both sides of the rotation curve, and we minimise the reduced chi squared statistic.

With \texttt{$^{\mathrm{3D}}$BAROLO}, there are nine possible parameters to fit: the rotation velocity ($V_{\mathrm{rot}}$), velocity dispersion ($\sigma$), radial velocity ($V_{\mathrm{rad}}$), systemic velocity ($V_{\mathrm{sys}}$), kinematic position angle ($PA_{\mathrm{kin}}$), kinematic centre ($x_{0,\mathrm{kin}},y_{0,\mathrm{kin}}$), $i$, and $Z_0$. To limit the number of free parameters, we leave only $V_{\mathrm{rot}}$, $\sigma$, and $PA_{\mathrm{kin}}$ of each ring free. For the remaining parameters, we fix the centre and inclination to the values returned by \texttt{CANNUBI}, and we fix $Z_0$ to zero following our finding that the thickness of the disc is unresolved ($< 710$ pc). We also fix $V_{\mathrm{rad}}$ to zero, meaning we assume no radial motion, following preliminary tests described below. For the systemic velocity, ($V_{\mathrm{sys}}$), we allow \texttt{$^{\mathrm{3D}}$BAROLO} to estimate this from the data cube, which results in $V_{\mathrm{sys}}=35.72$ km s$^{-1}$ from the [CII] line at a redshift $z=7.3065$, which is consistent with $V_{\mathrm{sys}}$ derived from a single component Gaussian fit to the [CII] spectrum and consistent with the reported uncertainty on the redshift from Table \ref{tab:R25 properties}. The results from the \texttt{$^{\mathrm{3D}}$BAROLO} fitting are given in Table \ref{tab:R25 parameters} and shown in Figure \ref{fig:barolo model}, \ref{fig:PVDs}, and \ref{fig:velocity curves}. We note that, whilst the $PA_{\mathrm{kin}}$ was allowed to vary for each ring, the best-fit model has a constant $PA_{\mathrm{kin}}$ across the disc.

To further test the robustness of our modelling, we also run \texttt{$^{\mathrm{3D}}$BAROLO} with all parameters left free. The resulting inclination ($i=32\pm5$) is consistent within the uncertainties with the $i$ fitted with \texttt{CANNUBI}, as is the fitted centre (within one beam FWHM of $x_{0,\mathrm{morph}},y_{0,\mathrm{morph}}$). The fitted values for $V_{\mathrm{rad}}$ are consistent with zero, within the uncertainties, for all five rings. The fitted $Z_0$ is also found to be consistent with zero. Whilst we adopt as our fiducial model the fit described above with only three free parameters ($V_{\mathrm{rot}}$, $\sigma$ and $PA_{\mathrm{kin}}$), we note that the overall results of this paper are consistent when the fit with all parameters left free is instead used.

\subsection{Results}
\label{sec:kinematic analysis}

\begin{figure*}
    \centering
    \includegraphics[width=0.9\textwidth]{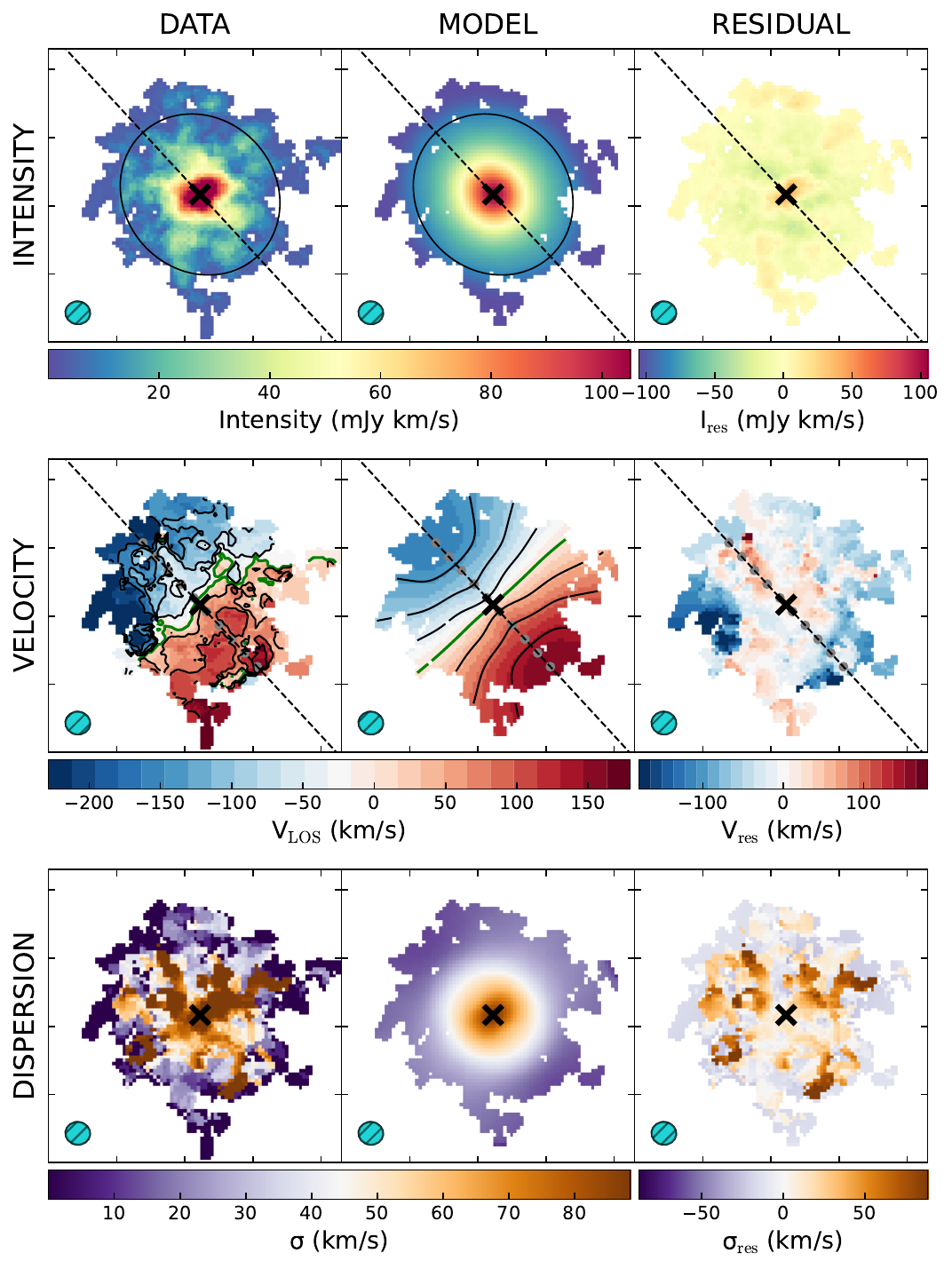}
    \caption { \texttt{$^{\mathrm{3D}}$BAROLO} fitting for REBELS-25. Emission is masked at 2$\sigma_{\mathrm{RMS}}$. The first column on the left shows the observed data, the middle column the model and the column on the right shows the residuals. The first row is for the intensity map, the second row for the velocity field map and the bottom row for the velocity dispersion map. In the first row, the black cross, ellipse and dashed line show the centre, radial extent and position angle of the \texttt{$^{\mathrm{3D}}$BAROLO} model, respectively. In the second row, the grey dots give an indication of the separation of each ring along the velocity field (0.11"). In the velocity field map of the data and model, we also plot the iso-contours from -180 to 180 km s$^{-1}$ in 45 km s$^{-1}$ increments. In all maps, the beam size is indicated by the turquoise ellipse in the bottom left corner.}
    \label{fig:barolo model}
\end{figure*}

\begin{figure}
    \centering
    \includegraphics[width=0.4\textwidth]{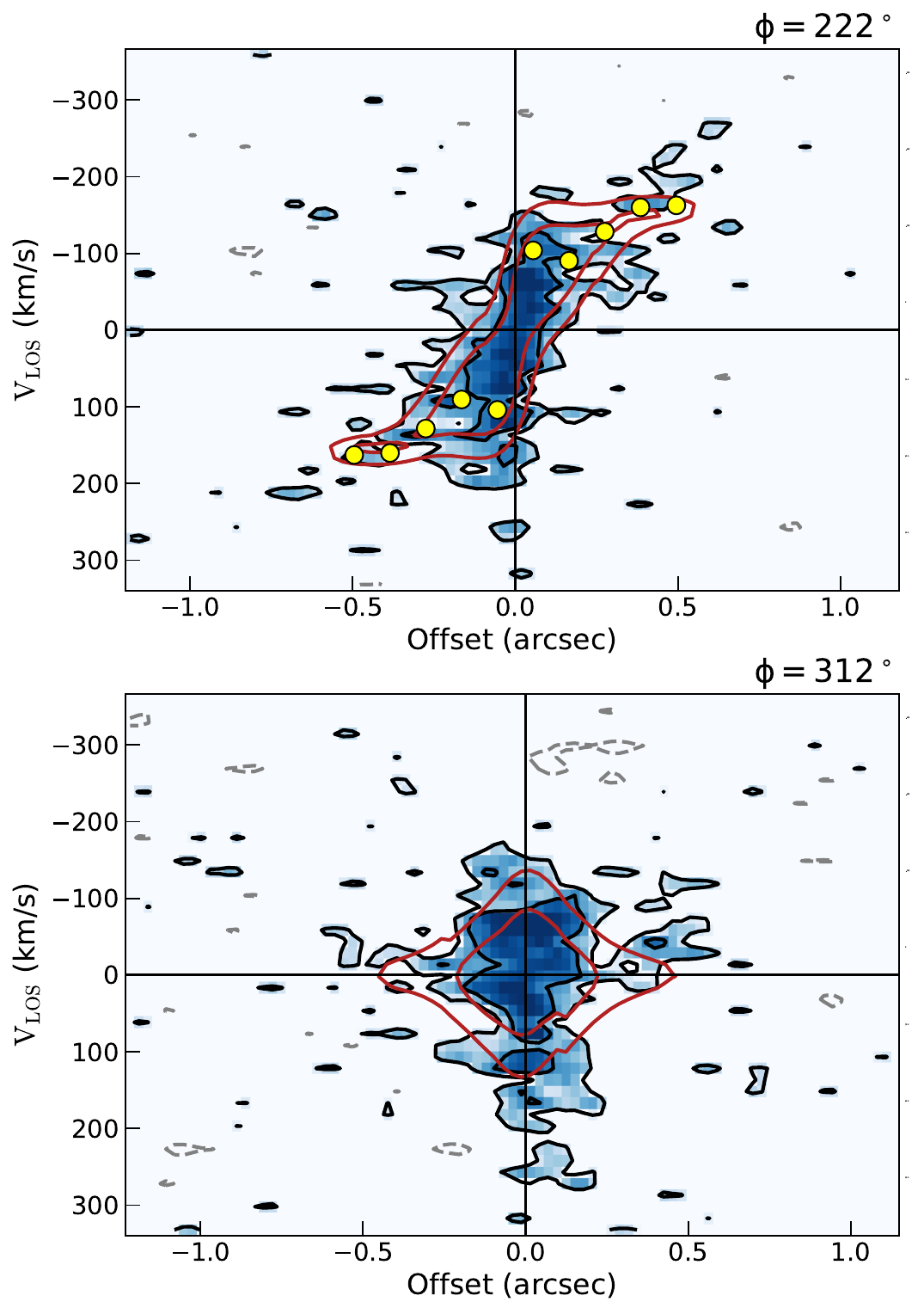}
    \caption {Major- (upper) and minor- (lower) axis position-velocity diagrams (PVDs) of REBELS-25, with black contours showing 2$\sigma_{\mathrm{RMS}}$ and 4$\sigma_{\mathrm{RMS}}$ emission, and the dashed grey contours -2$\sigma_{\mathrm{RMS}}$, where $\sigma_{\mathrm{RMS}}=168\mu$Jy/beam and is equal to one standard deviation above the median value as calculated by \texttt{$^{\mathrm{3D}}$BAROLO}. The red contours illustrate the best-fit model from \texttt{$^{\mathrm{3D}}$BAROLO}. In the major-axis PVD, the yellow markers show the rotation velocities of each ring projected along the line of sight.}
    \label{fig:PVDs}
\end{figure}

We show the final velocity maps from our fitting with \texttt{$^{\mathrm{3D}}$BAROLO} in Figure \ref{fig:barolo model} and the position velocity diagrams (PVDs) in Figure \ref{fig:PVDs}. We see a clear velocity gradient in the velocity field maps of Figure \ref{fig:barolo model} and a near-S-shape curve in the major axis PVD in Figure \ref{fig:PVDs}, which are characteristic features of rotating discs.

In Table \ref{tab:R25 parameters}, we report the final kinematic parameters derived by \texttt{$^{\mathrm{3D}}$BAROLO}, including the maximum rotation velocity ($V_{\mathrm{rot, max}}$), average velocity dispersion ($\bar{\sigma}$) and the $V_{\mathrm{rot, max}}/\bar{\sigma}$ ratio for REBELS-25. The $\bar{\sigma}$ gives a measure of the overall disc turbulence, and $V_{\mathrm{rot, max}}/\bar{\sigma}$ measures the ratio of ordered to turbulent motion.  For this analysis, we assume that the $V_{\mathrm{rot, max}}$ value is comparable to the rotational velocity at which the velocity curve flattens, since we see some flattening beyond 2 kpc in the rotation curve and major-axis PVD. 

We also plot the $V_{\mathrm{rot}}$ and $\sigma$ values of each ring as a function of radius in Figure \ref{fig:velocity curves}. We note that the rotation velocities reported are not corrected for pressure support, i.e., we have assumed that the rotation velocity we measure is an optimal tracer of the circular velocity, $V_{\mathrm{circ}}$. This is a reasonable assumption for REBELS-25 thanks to the high $V_{\mathrm{rot}}/\sigma$ values found, meaning that the pressure term in the Virial theorem is likely negligible (see e.g., Equations 6 and 7 in \citealt{iorio_little_2016}). We therefore take the $V_{\mathrm{rot, max}}$ value as the maximum circular velocity of the system in subsequent calculations.

The rotation velocity curve of REBELS-25 in Figure \ref{fig:velocity curves} shows the velocity rising quickly in the centre and then flattening to the maximum value of $\sim$370 km s$^{-1}$ at around 2 kpc. Similar rotation curves are observed in disk galaxies at lower redshifts (e.g., \citealt{noordermeer_mass_2007,lelli_sparc_2016, lelli_cold_2023,di_teodoro_flat_2016}). We also find a potential `bump' feature in the rotational velocity curve within the first $\sim$ 1 kpc. Whilst the uncertainties are such that this feature is not certain, this bump persists also in the fits with \texttt{$^{\mathrm{3D}}$BAROLO} where all the parameters are left free, and could indicate a potential bulge/bar component, as found in the Milky Way (e.g. \citealt{portail_dynamical_2017}), local galaxies (e.g. \citealt{lelli_sparc_2016}), and in some recent studies of galaxies at $z>4$ (e.g. \citealt{lelli_massive_2021, rizzo_dynamical_2021,lelli_cold_2023,tripodi_hyperion_2023, roman-oliveira_dynamical_2024}).

A decrease of $\sigma$ with radius is also observed, as found in some local disk galaxies (e.g., \citealt{bacchini_evidence_2020, fraternali_deep_2002, boomsma_hi_2008, iorio_little_2016}). This decrease is usually ascribed to the radial change of the mechanisms sustaining the turbulence within galaxies (e.g., supernova feedback, accretion and gravitational instabilities).

\subsubsection{Non-circular features in the \texttt{$^{\mathrm{3D}}$BAROLO} models}
\label{sec:barolo model issues}

Before drawing further conclusions based on the above modelling, we first investigate any features that are not well fit by \texttt{$^{\mathrm{3D}}$BAROLO}, which could indicate that the observed velocity gradient might be caused by outflows, interactions, or mergers, rather than solely the presence of a rotating disc (as discussed in e.g., \citealt{loiacono_multi-wavelength_2019,simons_distinguishing_2019}). Firstly, we see some evidence of twisted iso-velocity contours in the moment-1 map of REBELS-25 that are not well-fit by the rotating disc model with a constant $PA_{\mathrm{kin}}$ and $i$. These features could indicate a potential warped disc or distortions due to radial motions (e.g., \citealt{di_teodoro_radial_2021}). In some cases, twisted iso-velocity contours are also attributed to bars or other non-axisymmetric structures within galaxies (e.g., \citealt{wu_morphological_2021, buta_structure_1987, de_naray_kinematic_2009}). Secondly, in the velocity field map and the residual map of Figure \ref{fig:barolo model} we can see a region on the approaching side (to the West) of the galaxy with high velocities which is not reproduced within the model. This feature can also be seen in the channel maps at $-228$ and $-274$ km s$^{-1}$ (Figure \ref{fig:channel maps}). This could have a number of causes, such as recently accreted gas, streaming motions along spiral arms, or due to a warped gas disc. These features may also be consistent with noise in the data. In an attempt to investigate these features further, we tested \texttt{$^{\mathrm{3D}}$BAROLO} fits that incorporate radial motions. As mentioned in Section \ref{sec:kinematic modelling set up}, we find $V_{\mathrm{rad}}$ consistent with zero given the uncertainties, and we also find that these models were not able to reproduce similar high velocity regions. 

Additionally, we see some faint emission beyond the rotating disc model in the PVDs. For example, in the lower half of the minor-axis PVD, and also in the lower half of the major-axis PVD, we see some faint high receding velocity regions. As with the features described above, these could signify non-circular motions, such as vertical/radial motions due to inflows, outflows or mergers, the presence of a bar or the presence of spiral arms. Similar deviations due to non-circular motions are common in rotating galaxies at $z = 0$ (e.g., \citealt{fraternali_new_2001, jorsater_high_1995, zurita_ionized_2004, trachternach_dynamical_2008, erroz-ferrer_h-alpha_2015}), and similar faint emission outside ideal rotating disc models can also be seen in mock data of simulated rotating discs (e.g., see Figure 6 of \citealt{rizzo_dynamical_2022}). Distinguishing between the different processes potentially responsible for these features is not feasible at the current resolution and SNR, and we note that this emission is only detected at $\sim 2\sigma_{\mathrm{RMS}}$.

Overall, we find that the majority of the emission is well described by the rotating disc model, although we cannot exclude some disturbed kinematics due to, for example, outflows or a warped disc. Follow-up data on the stellar morphology would be needed to confirm the presence of any non-axisymmetric features that could also be introducing non-circular motions.

\subsubsection{Secondary [CII] component}
\label{sec:secondary component}

In \cite{hygate_alma_2023}, a potential secondary [CII] component is identified in the range $\sim +250$ to $+650$ km s$^{-1}$ in the [CII] spectrum of REBELS-25 from the lower resolution LP data. Within \cite{hygate_alma_2023}, it is hypothesised that this spectral component could be the result of an outflow (where it is modelled as a broad component with width $\sim 380$ km s$^{-1}$) or a minor merger (modelled as a narrow component with width $\sim 100$ km s$^{-1}$).  We have searched for this secondary [CII] component in our data set following a similar methodology as in \cite{hygate_alma_2023}, whereby we extracted a spectrum for REBELS-25 using a circular aperture of 1.75" and attempted to fit a main and secondary [CII] component, where we limit the centre of the secondary peak to the same limits used in \cite{hygate_alma_2023} for both the outflow and minor merger model. However, we find no convincing secondary component in the [CII] spectrum from the high resolution data. We also extracted spectra from beam sized apertures across the source in case the component has been diluted in the global spectrum, as found for a different galaxy by \cite{herrera-camus_kiloparsec_2021}, but again find no significant secondary component. Indeed, \cite{hygate_alma_2023} suggested the component might instead be extended on $\sim9$ kpc scales. This would make the large-scale outflow scenario the more likely option. Investigating the nature of this potential component further would therefore require deeper data with increased surface brightness sensitivity to such extended structure.

\begin{table}
\caption[]{Table showing the [CII] morphological and kinematic parameters derived from a 2D Sérsic model, \texttt{CANNUBI} and \texttt{$^{\mathrm{3D}}$BAROLO}, as described in the text.}
\normalsize
\centering
\begin{tabular}{lll}
\centering
        
        Parameter & Value & Model used \\
	\hline\hline
	$n$ & $1.3 \pm 0.2$& 2D Sérsic model \\
    $r_e$ (kpc) & $2.2 \pm 0.3$ & 2D Sérsic model\\
    $PA_{\mathrm{morph}}$ ($^{\circ}$) & $215 ^{+12}_{-31}$ & \texttt{CANNUBI}\\
    $\mathrm{PA}_{\mathrm{kin}}$ ($^{\circ}$) & $222 \pm 5$ & \texttt{$^{\mathrm{3D}}$BAROLO}\\
    $i$ ($^{\circ}$) & $25\pm 6$& \texttt{CANNUBI}\\
    $V_{\mathrm{rot, max}}$ (km s$^{-1}$) &  $372^{+82}_{-66}$& \texttt{$^{\mathrm{3D}}$BAROLO}\\
    $\bar{\sigma}$ (km s$^{-1}$) & $33 \pm 9$ & \texttt{$^{\mathrm{3D}}$BAROLO}\\
    $V_{\mathrm{rot, max}}/\bar{\sigma}$ & $11^{+8}_{-4}$ & \texttt{$^{\mathrm{3D}}$BAROLO}\\
        \hline\hline
	
\end{tabular}

\label{tab:R25 parameters}
\end{table}

\subsubsection{Merger or rotating disc?}
\label{sec:merger or disc?}

In some cases at low spatial resolution, the observed velocity map of a merging system can be similar to that of a smooth rotating disk, as observations can smooth out the irregularities and asymmetries of a merger (e.g., \citealt{simons_distinguishing_2019, kohandel_velocity_2020, rizzo_dynamical_2022}). Within the literature, there exist multiple methods and criteria used for differentiating between rotating discs and mergers. Here we describe the results from applying the rotating disc criteria laid out in \cite{wisnioski_kmos3d_2015}, and then adapted for [CII] ALMA-ALPINE data in \cite{jones_alpine-alma_2021}, and also the PV Split method described in \cite{rizzo_dynamical_2022}.

\begin{figure}
    \centering
    \includegraphics[width=0.48\textwidth]{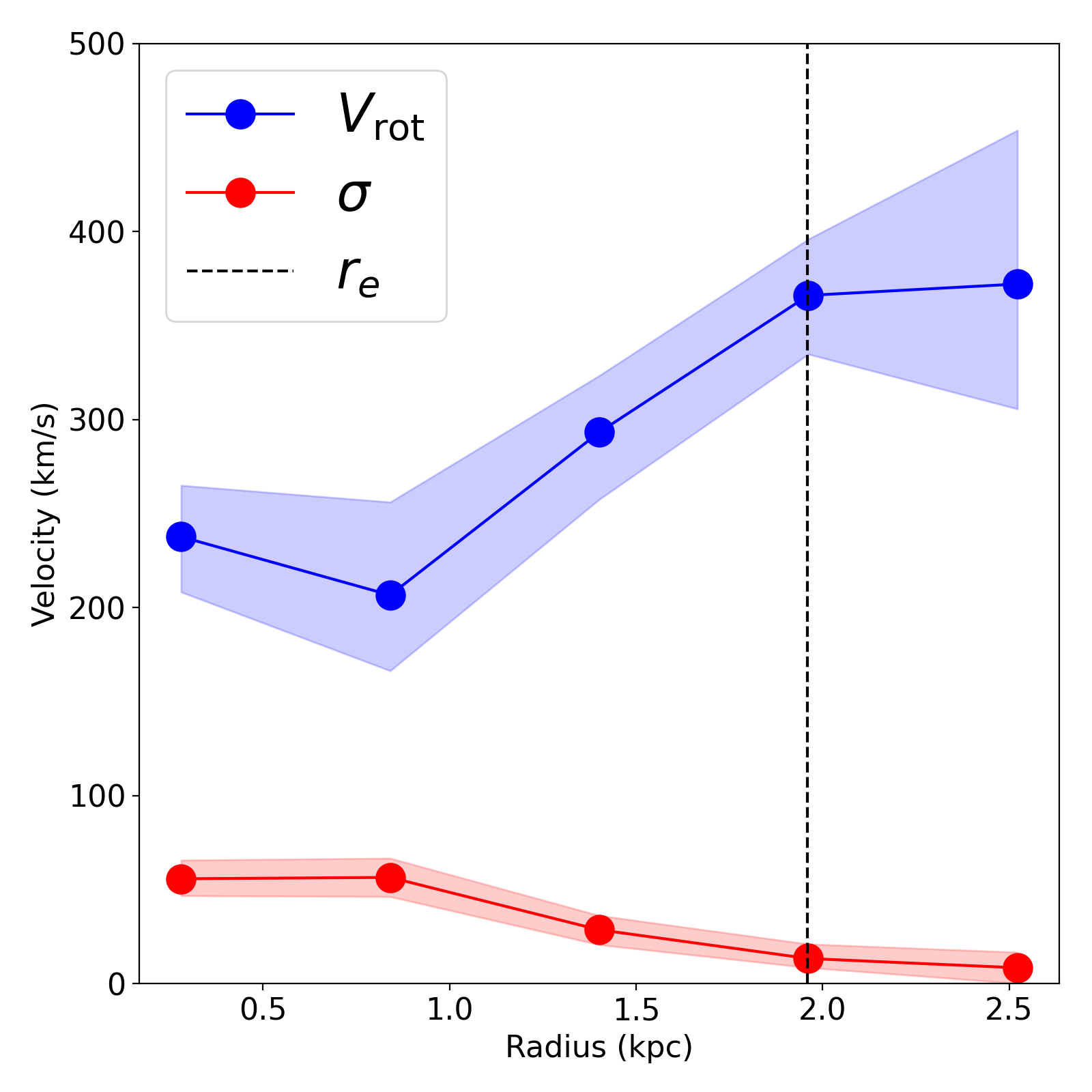}
    \caption {The rotation velocity ($V_{\mathrm{rot}}$, dark blue) and velocity dispersion ($\sigma$, red) curves for REBELS-25. The size of the band indicates the estimated uncertainty from the \texttt{$^{\mathrm{3D}}$BAROLO} fit for $V_{\mathrm{rot}}$ and $\sigma$. The dashed black line indicates the effective radius from \texttt{Sersic2D} fitting.}
    \label{fig:velocity curves}
\end{figure}

We use the same five rotating disc criteria listed in \cite{jones_alpine-alma_2021} with the 2D intensity, rotational velocity and velocity dispersion maps outputted from \texttt{$^{\mathrm{3D}}$BAROLO}. We also make use of the morphological parameters fitted with \texttt{CANNUBI}, and we use  \texttt{$^{\mathrm{3D}}$BAROLO}'s 2D fit capabilities to an inputted velocity field for the third and fifth criteria.

In \cite{jones_alpine-alma_2021}, typically at least 3 of these criteria are met for the ALPINE galaxies classified as rotating discs. For REBELS-25, we find that all 5 criteria are met:

\begin{enumerate}
    \item We find that a slice along the kinematic major axis reveals a slope in position-velocity space with $>3\sigma$ significance, implying a smooth velocity gradient.
    \item For each of the five fitted rings, we find $V_{\mathrm{rot}}>\sigma$ (see Figure \ref{fig:velocity curves}).
    \item We find that the kinematic centre derived by a 2D fit with \texttt{$^{\mathrm{3D}}$BAROLO} and the peak in the velocity dispersion map are consistent within the uncertainties.
    \item The $PA_{\mathrm{morph}}$ from \texttt{CANNUBI} and the average $PA_{\mathrm{kin}}$ across all rings are consistent within the uncertainties, as expected for an ideal rotating disc.
    \item We find that the kinematic centre fitted by \texttt{$^{\mathrm{3D}}$BAROLO} and the morphological centre fitted by \texttt{CANNUBI} are consistent within error.
\end{enumerate}

From these tests, we may conclude that REBELS-25 is a rotating disc. However, \cite{rizzo_dynamical_2022} find that these criteria can mistakenly classify galaxy kinematics, and we therefore also use the PV Split method described in the same paper.

In \cite{rizzo_dynamical_2022}, mock ALMA data cubes of SERRA simulated rotating discs, disturbed discs, and merging galaxies (\citealt{pallottini_survey_2022}) were used to show that the morphology and symmetry of the major and minor axis PVDs can be used to kinematically classify galaxies, with rotating discs typically showing a symmetric, S-shaped position-velocity profile. Three parameters are defined to measure the morphology and symmetry of the PVDs, and a plane that divides these two kinematic classes in the  3D PV Split parameter space is defined and tested using local disc galaxies and mergers from the WHISP survey (\citealt{van_der_hulst_westerbork_2001-1}) in \cite{roman-oliveira_regular_2023}. Using the PVDs returned by \texttt{$^{\mathrm{3D}}$BAROLO} for REBELS-25, we find that it lies in the rotating disc parameter space of the 3D PV Split diagram. However, we again cannot exclude a disturbed disc or minor merger scenario with this method.

As described above, we have found some evidence of non-circular features and therefore cannot rule out the existence of a minor merging component or outflows/inflows with the current data. However, following visual inspection of the \texttt{$^{\mathrm{3D}}$BAROLO} fits, tests using the rotating disc criteria of \cite{wisnioski_kmos3d_2015}, and tests using the PV Split method of \cite{rizzo_dynamical_2022}, we find that REBELS-25 is best described as a rotation-dominated disc galaxy.

\subsubsection{Dynamical mass and gas mass}
\label{sec:dynamical mass}

For a system in virial equilibrium, the dynamical mass, $M_{\mathrm{dyn}}$, enclosed within a radius, $R$, is given by:

\begin{equation}
\label{eq:dynamical mass}
M_{\mathrm{dyn}}(<R) =\frac{{k(R)} R v_{\mathrm{circ}}^2(R)}{G},
\end{equation}

\noindent where $k(R)$ is the virial coefficient at radius $R$. \cite{price_kinematics_2022} use Sérsic models to derive the virial coefficients for a range of $n$ and intrinsic axis ratios ($q$). As we have some evidence that REBELS-25 is a thin exponential disc from \texttt{CANNUBI} and 2D Sérsic fits, we adopt a total virial coefficient (which relates the total dynamical mass of the system to the circular velocity) of $k_{\mathrm{tot}}=1.8$, which is the virial coefficient for $n=1$ and $q=0.2$ (typically quoted for thin discs, e.g. \citealt{van_der_wel_geometry_2014}). We take the $V_\mathrm{rot}$ in the 4th ring ($=366^{+30}_{-31}$ km s$^{-1}$), since this is the closest to the $r_e$ as derived from the 2D Sérsic fitting. This results in $M_{\mathrm{dyn, tot}} = (1.2\pm0.3) \times 10^{11} \mathrm{ M_{\odot}}$. This is larger but within the uncertainties of the value derived in \cite{hygate_alma_2023}, which used the low resolution LP data.

Assuming that the contribution of dark matter to the mass budget within $r_e$ is negligible (as found at high-$z$ in e.g., \citealt{van_dokkum_forming_2015, price_mosdef_2016, wuyts_kmos3d_2016, genzel_strongly_2017, gelli_stellar_2020}), we can estimate the total gas mass from $M_{\mathrm{gas}}=M_{\mathrm{dyn}}-M_*$. Adopting a stellar mass of $M_* = 8^{+4}_{-2}\times10^9 \mathrm{ M_{\odot}}$ results in $M_{\mathrm{gas}}=(1.1^{+0.4}_{-0.3})\times 10^{11} \mathrm{M_{\odot}}$, with a gas fraction of $f_{\mathrm{gas}} = M_{\mathrm{gas,tot}}/(M_{\mathrm{gas,tot}}+M_*) = 0.93^{+0.03}_{-0.07}$. However, we note that if we adopt the stellar mass derived in \cite{topping_alma_2022} using a non-parametric star formation history ($M_*=19^{+5}_{-8} \times 10^9 \mathrm{ M_{\odot}}$), $M_{\mathrm{gas,tot}}=(9.8^{+4.3}_{-3.5})\times 10^{10} \mathrm{ M_{\odot}}$ and $f_{\mathrm{gas}} = 0.84^{0.09}_{-0.11}$.

With these constraints on the gas mass, we can also derive a conversion factor between the [CII] luminosity and $M_{\mathrm{gas}}$, $\alpha_{\mathrm{[CII]}}$, assuming that the [CII] emission is a good tracer of the total gas mass (\citealt{swinbank_alma_2012,gullberg_dust_2018}; \citealt{zanella_c_2018}). With a double Gaussian fit to the [CII] spectrum, we find $L_{\mathrm{[CII]}}=(1.8\pm0.2) \times 10^9 L_{\odot}$, which is consistent within the uncertainties with the value derived from the low resolution LP data in \cite{hygate_alma_2023}. This results in $\alpha_{\mathrm{[CII]}}=62^{+32}_{-22} \mathrm{M_{\odot}/L_{\odot}}$ (or $55^{+35}_{-23}$ if the \citealt{topping_alma_2022} $M_*$ is used). These values are a factor of $\sim 10$ times higher than the median value found for dusty star-forming galaxies at $z~4.5$ in \cite{rizzo_dynamical_2021}, and a factor of $\sim$ 2 times higher than the value of 30 $\mathrm{M_{\odot}/L_{\odot}}$ found for main sequence galaxies in \cite{zanella_c_2018} (although we note that this conversion is derived for the molecular gas mass, and not total gas mass). However, we note that the uncertainties are high, and a more statistically significant sample would be necessary to investigate $\alpha_{\mathrm{[CII]}}$ values for star forming galaxies at $z>6$.

From the properties derived by \texttt{$^{\mathrm{3D}}$BAROLO}, and an estimate for the gas mass, it is also possible to determine the Toomre parameter (\citealt{toomre_gravitational_1964}), $Q$, of REBELS-25. This $Q$ parameter is used to measure the local instability of a disc, with $Q<1$ indicating that it is susceptible to local gravitational perturbations (LGIs) (e.g., \citealt{leung_dynamical_2020}). Using Equation (1) from \cite{neeleman_cold_2020}, we find an average $Q$ of $0.4^{+0.5}_{-0.2}$ over the radial extent of the [CII] emission, suggesting it may be locally unstable. However, some studies at $z\sim0$ have found that $Q$ does not show any correlation with fragmentation or star formation (e.g  \citealt{leroy_star_2008, romeo_simple_2013, elmegreen_star_2015, romeo_specific_2023}), and a recent study also indicates that high-$z$ galaxies are not as locally unstable as expected from the $Q$ parameter (\citealt{bacchini_3d_2024}).

\section{Discussion}
\label{sec:discussions}

Through our kinematic analysis of REBELS-25, we have found a dynamically cold disc with a high degree of rotational support ($V_{\mathrm{rot, max}}/\bar{\sigma}=11^{8}_{-4}$) and low overall turbulence ($\bar\sigma=33\pm9$ km s$^{-1}$). In \cite{kohandel_dynamically_2023}, galaxy discs are classed according to their $V/\sigma$ ratios, with $V/\sigma>10$ being defined as `super cold' discs. Following this classification, we find that REBELS-25 is the most distant super cold disc galaxy observed to date. In the following, we first discuss our findings on this galaxy in more detail (Section \ref{sec:discussion1}) before placing REBELS-25 in context with other high resolution gas kinematic studies.

\subsection{An evolved, dynamically cold rotating disc at $z=7.31$}
\label{sec:discussion1}

In Section \ref{sec: cii morphology}, we found that the [CII] emission of REBELS-25 is well-fit by an exponential disc model, similar to what is found for galaxies in the local Universe and for the dust continuum in some submillimetre galaxies at $z\sim1-3$ (e.g., \citealt{hodge_kiloparsec-scale_2016, barro_sub-kpc_2016, rivera_resolving_2018, fujimoto_alma_2018, gullberg_alma_2019}). This is in contrast to the clumpy and irregular morphologies observed and anticipated at $z>6$ (see Section \ref{sec:intro}). We also find that the gas disc of REBELS-25 is likely thinner than the $Z_0 \sim 1$ kpc derived from galaxies at $z\sim4.5$ in \cite{roman-oliveira_regular_2023}. This contradicts the general expectation that galaxies at higher redshifts will tend to have thicker discs due to increased turbulence, as found in $\Lambda$CDM galaxy formation simulations (e.g., \citealt{brook_thin_2012}; \citealt{bird_inside_2013, park_new_2019, bird_inside_2021, renaud_vintergatan_2021}).

We also see some tentative evidence that REBELS-25 is a barred disc (Section \ref{sec:potential features}), with a sudden change in both the position angle and ellipticity of the [CII] and dust continuum emission at $\sim 1$ kpc. With the current data, it is not feasible to search for kinematic signatures of a bar, and we therefore cannot rule out alternative scenarios, such as inflows/outflows or a late-stage minor merger. However, a bar may be feasible since barred structures are thought to form relatively quickly (order of a hundred million years) in massive dynamically cold disc galaxies (e.g., \citealt{hohl_numerical_1971, sellwood_dynamics_1993, athanassoula_bar-halo_2002, rosas-guevara_evolution_2022, bland-hawthorn_rapid_2023}). Barred galaxies are common in the nearby Universe (e.g., \citealt{knapen_subarcsecond_2000, whyte_morphological_2002, laurikainen_comparison_2004, buta_classical_2015}), but the fraction of barred galaxies has been found to decrease between $z =$ 0 to 1 (e.g., \citealt{sheth_evolution_2008}). However, some recent observations have identified barred galaxies at $z=1-4.4$ (\citealt{simmons_galaxy_2014,   hodge_alma_2019, tsukui_detecting_2023, huang_j0107a_2023, smail_hidden_2023, guo_first_2023,amvrosiadis_onset_2024, leconte_jwst_2024}), further indicating that such structures can form in galaxies within the first few Gyrs after the Big Bang.

We additionally see some very tentative evidence of arm-like or clump-like extended features at $\sim$ 3 kpc to the North and South of the peak of emission in the dust continuum map (Figure \ref{fig:presenting data}). Similar features have been observed in the dust emission of SMGs (e.g., \citealt{gullberg_alma_2019, hodge_alma_2019}). However, these features are very faint. Three approved JWST programs (GO-1626, GO-6036 and GO-6480) targeting this galaxy and providing both low resolution spectroscopy and high spatial resolution imaging with grism spectroscopy of the stellar and ionised gas content may shed more light on whether or not any bar, spiral arm and/or clump structures are present. 

Some clumpy structure has already been identified in REBELS-25 from the rest-frame UV imaging with HST. As discussed in \cite{hygate_alma_2023}, these UV clumps could have a number of explanations, including differential dust obscuration, regions of intense star formation or potential merger activity. From the analysis described thus far, the high resolution [CII] observations disfavour the hypothesis that these clumps are the result of an ongoing merger, since we see no strong kinematic evidence of merging activity. Our analysis instead indicates that these clumps may be due to differential dust obscuration, particularly at the centre of the disc where the dust emission peaks, or due to star forming clumps. Similar clumpy morphology in the rest-frame UV embedded within a rotating gas disc has also been found in, for example, a $z=6.072$ galaxy (\citealt{fujimoto_primordial_2024}).

We also find that REBELS-25's dynamical mass is a factor of $\sim 14$ times greater than the adopted stellar mass from Table \ref{tab:R25 properties}, where this stellar mass is obtained with the \texttt{BEAGLE} SED modelling code assuming a constant star formation history (SFH) and a \cite{chabrier_galactic_2003} IMF, with full details described in Stefanon et al. (in prep). If we instead adopt the stellar mass derived in \cite{topping_alma_2022}($M_*=19^{+5}_{-8} \times 10^9 /M_{\odot}$), $M_{\mathrm{dyn}}/M_*$ becomes $\sim 6$. \cite{de_graaff_ionised_2023} have recently reported $M_{\mathrm{dyn}}/M_*$ values as high as 30 in high-$z$ galaxies, whereas studies of $z\sim2$ galaxies find $M_{\mathrm{dyn}}/M_* \sim 4$ (\citealt{maseda_confirmation_2013}). For the case of REBELS-25, we have seen that the gas fraction is likely very high, however we cannot yet rule out contributions from dark matter components to the mass budget.

\subsection{Selecting a comparison sample}

\begin{figure*}
    \centering
    \begin{subfigure}[b]{0.32\textwidth}
        \centering
        \includegraphics[width=\textwidth]{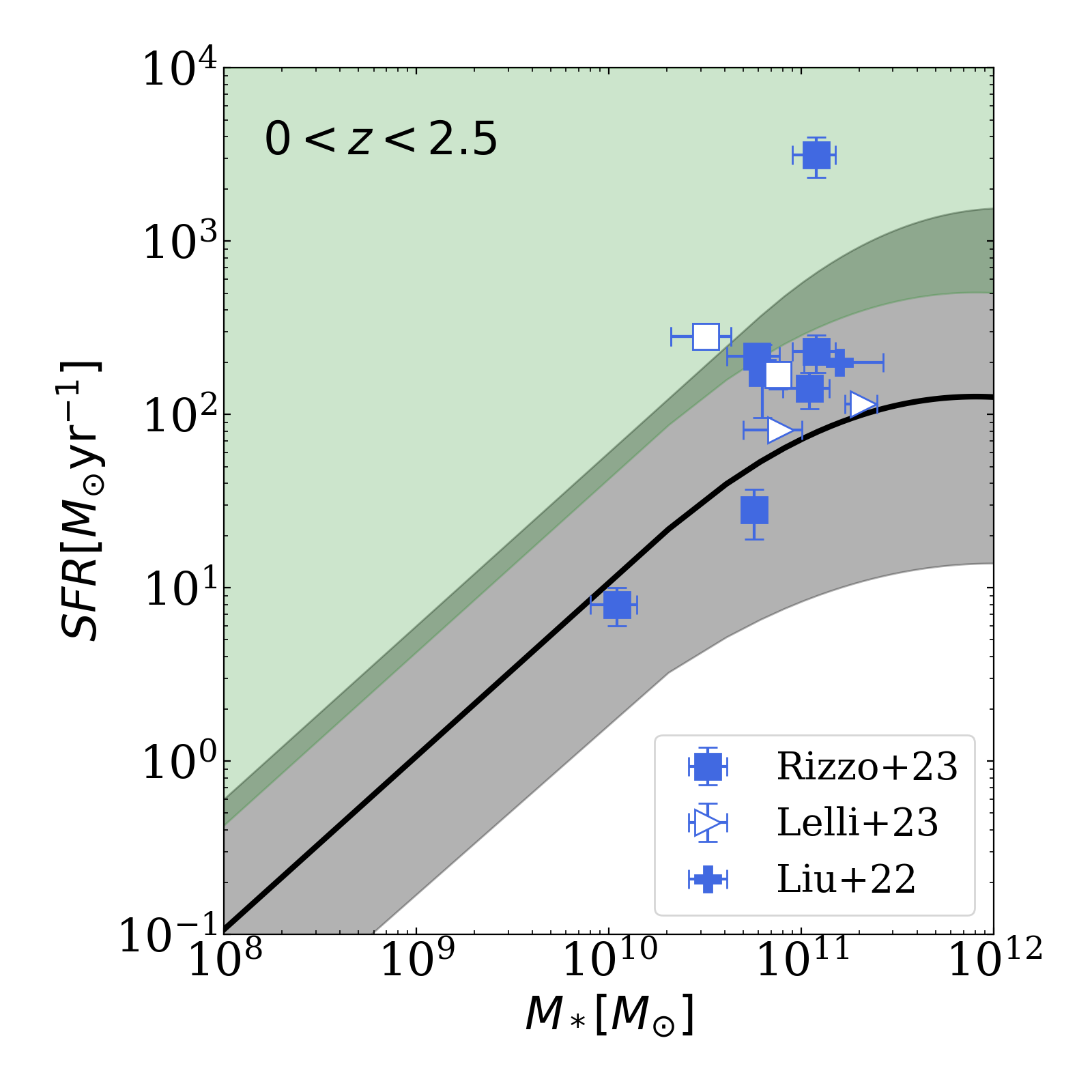}
        \caption{}
        \label{fig:main sequence 1}
    \end{subfigure}
    \begin{subfigure}[b]{0.32\textwidth}
       \centering
        \includegraphics[width=\textwidth]{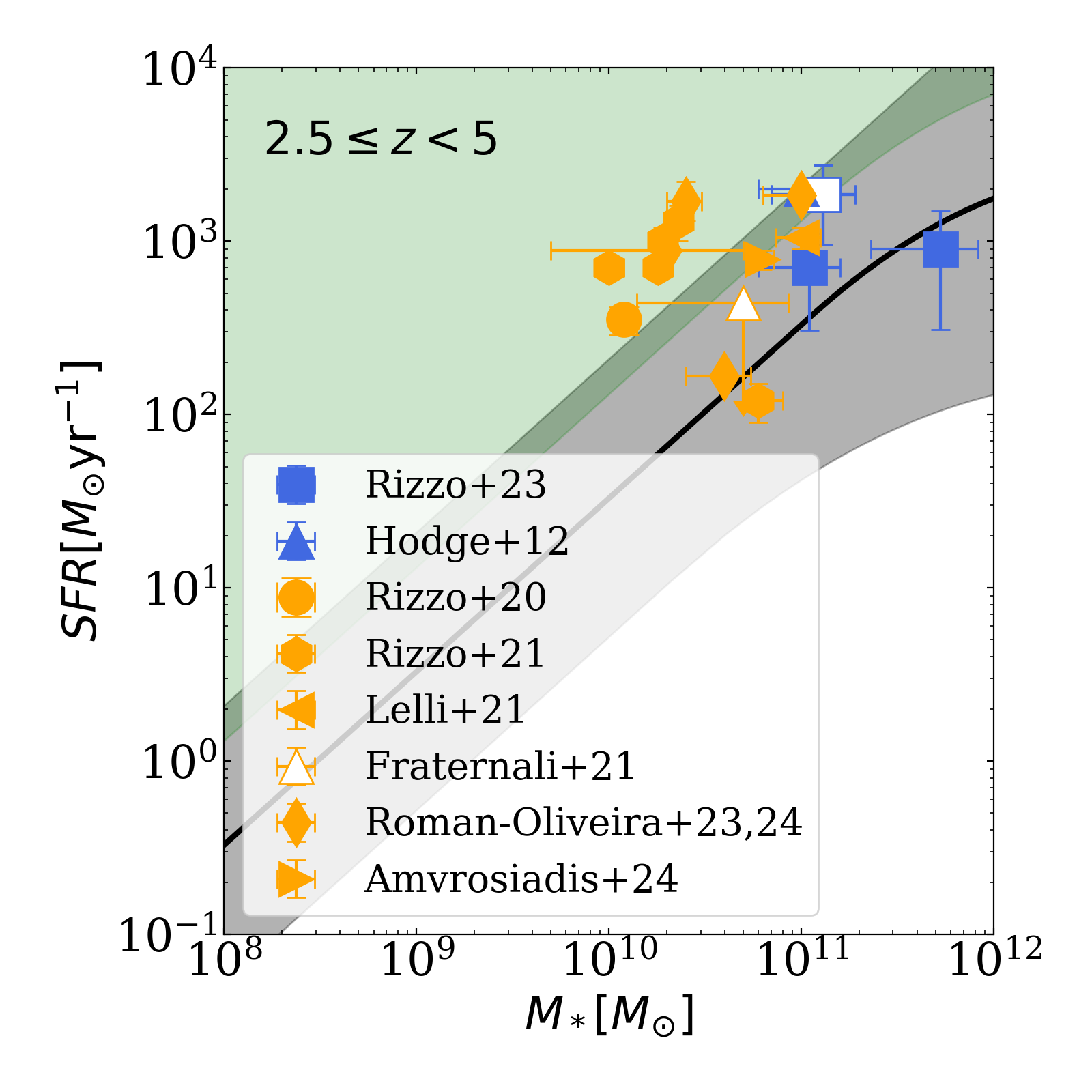}
        \caption{}
        \label{fig:main sequence 2}
    \end{subfigure}
    \begin{subfigure}[b]{0.32\textwidth}
       \centering
        \includegraphics[width=\textwidth]{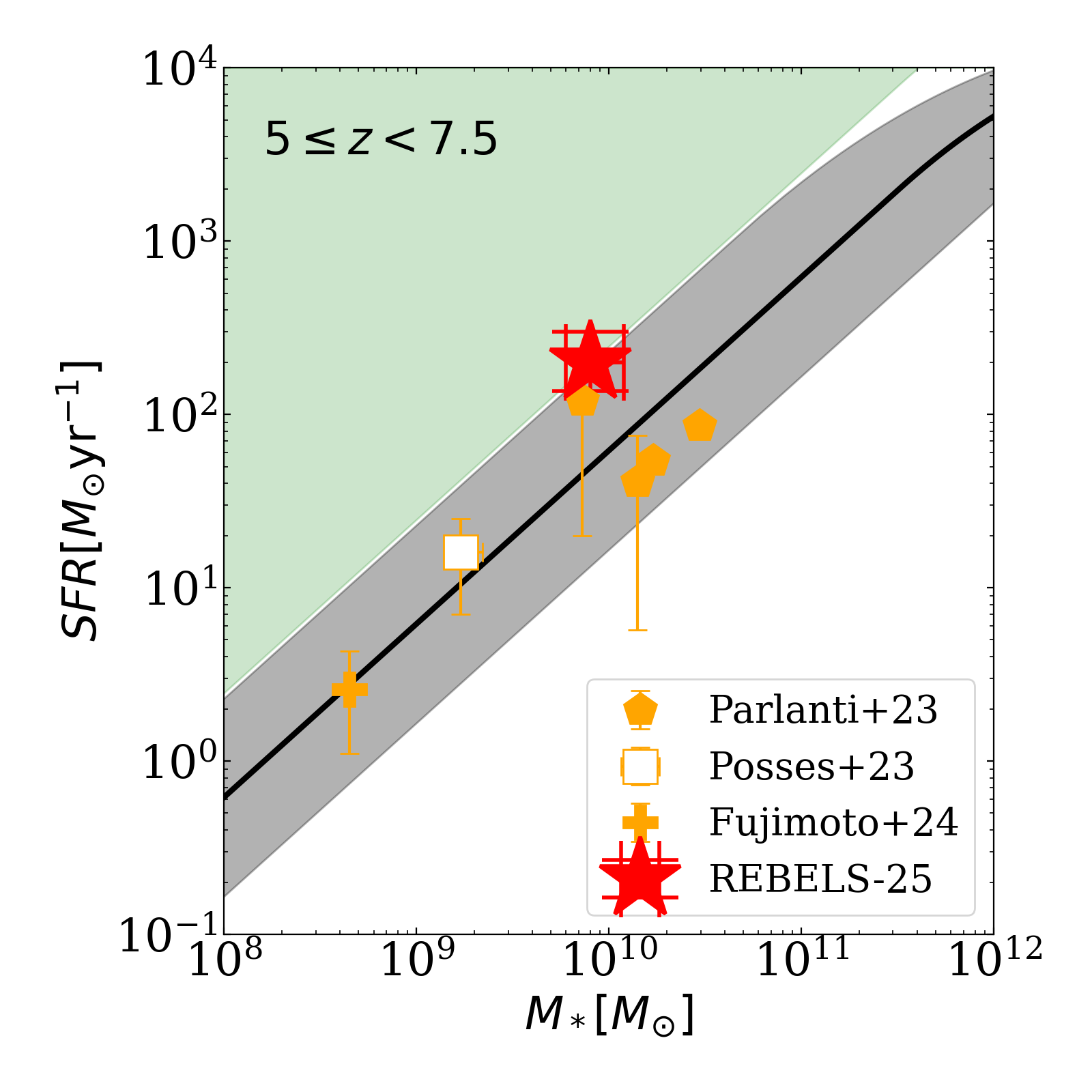}
        \caption{}
        \label{fig:main sequence 3}
   \end{subfigure}
    
    \caption {Distribution of REBELS-25 and a few comparison samples in the $M_*$ - SFR plane. We split the sample into three redshift bins and show the main-sequence line from \protect\cite{schreiber_herschel_2015} for $z\sim$ 1.25, 3.75 and 6.25 in panels (a), (b) and (c), respectively, by solid black lines. The grey shaded areas show the 1$\sigma$ uncertainty on the main-sequence lines, and the green shaded areas show the region occupied by starbursts according to a threshold of $4 \times$ the main-sequence lines (e.g. \protect\citealt{rodighiero_lesser_2011}). Galaxies observed with [CII] emission are plotted with orange markers (apart from REBELS-25, which is plotted with a large red star-shaped marker in panel c), whilst galaxies observed with either [CI] or CO are shown with blue markers. Empty markers indicate galaxies where the kinematic properties are given as limits only.}
    \label{fig:comparison sample}
\end{figure*}

To place REBELS-25 into context with the evolution of the gas kinematic properties of galaxies across cosmic time, a comparison sample across a broad redshift range with comparable observations is necessary. One important point to consider when comparing kinematic studies at a range of redshifts is the ISM tracer used. At $z\sim0$, gas kinematics of galaxies are typically traced by warm ionised gas tracers, such as H$\alpha$. These warm ionised gas tracers have been found to result in higher velocity dispersion compared to values derived from atomic or molecular gas (e.g., \citealt{levy_edge-califa_2018, ubler_evolution_2019, girard_systematic_2021, kohandel_dynamically_2023}). We therefore mainly compare with studies using cold gas tracers, and we discuss the implications of any comparisons made with different tracers in subsequent sections. 

As discussed in Section \ref{sec:intro}, the [CII] 157.7$\mu$m emission line is a useful, strong-line tracer of cold gas kinematics. However, observations of this line become increasingly difficult at $z \lesssim 3.5$ from ground-based instruments due to the poor transmission through Earth's atmosphere. For this reason, we compile literature results of both high resolution [CII] kinematics studies, typically at $z > 3.5$, and high resolution [CI] and CO kinematics studies, typically at $0.5 \lesssim z \lesssim 3.5$. For this selection from the literature, we take into consideration the criteria recommended in \cite{rizzo_dynamical_2021} for distinguishing between mergers and rotating discs. We therefore select only literature studies that use observations with $\gtrsim 3$ independent resolution elements across the semi-major axis based on the reported radial extent of emission of the galaxies and the beam size of their corresponding observations. Our final literature compilation consists only of galaxies that are classified as rotating discs based on these high-quality kinematic data. We discuss the literature sample in more detail below.

\cite{rizzo_alma-alpaka_2023} presented the ALMA-ALPAKA survey; a collection of 28 star forming galaxies with high resolution, high SNR ALMA archival observations of CO and [CI] emission at $z = 0.5 - 3.5$. The physical resolution of these sources varies from 1 to 4 kpc. Whilst sub-kpc resolution observations are preferable to robustly analyse kinematics, this resolution is found to be sufficient to sample the velocity curves with $\gtrsim 3$ resolution elements for 13 discs (classified as rotating discs using the PV Split method) within the sample. Of these 13 discs within the ALMA-ALPAKA survey, seven are classified as starbursts, and six as main sequence galaxies (the remaining galaxy has no constraint on $M_*$ or SFR). Additionally, five of these are classified as AGN galaxies or AGN candidates by \cite{rizzo_alma-alpaka_2023}. Many of these 13 discs were also the focus of previous kinematic studies, as described in \cite{rizzo_alma-alpaka_2023}.

\cite{parlanti_alma_2023} collected and analysed all available Band 4, 5, and 6 ALMA archival observations of [CII] (and [OIII]88$\mu$m) emission in galaxies at $z > 3.5$ with angular resolution $<$ 1.5" and SNR $>$ 7. From this sample, we initially select seven galaxies that meet our criteria. Of these seven galaxies, five were studied using the same data set in other works. These five galaxies are DLA0817g, ALESS 073.1, HZ4, HZ7, and COS-29, analysed previously by \cite{neeleman_cold_2020} (and \citealt{roman-oliveira_regular_2023}), \cite{lelli_massive_2021, herrera-camus_kiloparsec_2022, lambert_extended_2022}, and \citealt{posses_structure_2023}, respectively. \cite{lelli_massive_2021}, \cite{posses_structure_2023}, and \cite{roman-oliveira_regular_2023} use \texttt{$^{\mathrm{3D}}$BAROLO} for their kinematic analyses, and \cite{neeleman_cold_2020} use \texttt{QUBEFIT} but find that \texttt{$^{\mathrm{3D}}$BAROLO} produces consistent results. To be more comparable to our analysis of REBELS-25 and the kinematic analyses of the ALMA-ALPAKA galaxies in \cite{rizzo_alma-alpaka_2023}, who also use \texttt{$^{\mathrm{3D}}$BAROLO}, we therefore use the kinematic properties derived from \cite{roman-oliveira_dynamical_2024},  \cite{lelli_massive_2021}, and \cite{posses_structure_2023} for DLA0817g, ALESS 073.1, and COS-29, respectively. The galaxy HZ7 is found to be a potential merger by the kinematic analysis of \cite{lambert_extended_2022}, and we therefore exclude it from our selection. In \cite{herrera-camus_kiloparsec_2022}, the 3D parametric code DYSMAL is used to model the kinematics of HZ4. The resulting kinematic properties derived in \cite{herrera-camus_kiloparsec_2022} and \cite{parlanti_alma_2023} are consistent for HZ4, and therefore for HZ4 and the remaining two galaxies (HZ9, J1211, and VR7) selected from \cite{parlanti_alma_2023}, we use the kinematic properties derived therein.

To increase the size of our comparison sample, we also consider starburst, quasar host, sub-millimetre, and gravitationally lensed galaxies that do not show signs of merging activity from \cite{roman-oliveira_regular_2023, rizzo_dynamically_2020, liu_600_2023, fraternali_fast_2021, hodge_evidence_2012,fujimoto_primordial_2024-1, neeleman_alma_2023,lelli_cold_2023}; and \cite{amvrosiadis_onset_2024}. There exist additional studies of cold gas kinematics (including HI kinematics) with observations of sufficient angular resolution and sensitivity, particularly at lower redshifts, however we choose to focus on higher redshifts ($z>0.5$) where observations are more comparable to those of REBELS-25. Additional galaxies considered but not included in the comparison sample include AZTEC-1, a sub-millimetre galaxy at $z=4.6$ with high resolution [CII] ALMA observations initially analysed by \cite{tadaki_gravitationally_2018}, who found a rotating disc with $V_{\mathrm{rot, max}} \sim 227$ km s$^{-1}$ and $\bar{\sigma} \sim 74$ km s$^{-1}$. However, when the same dataset was analysed in \cite{roman-oliveira_regular_2023}, they found evidence of merging activity according to the disc criteria of \cite{wisnioski_kmos3d_2015}, the PV Split method, and from a visual inspection of the velocity field and PVDs. \cite{roman-oliveira_regular_2023} therefore conclude that AZTEC-1 is likely a merger. Other notable high redshift cold gas kinematic studies include \cite{rivera_resolving_2018, rybak_strong_2019, tamura_300_2023, pope_alma_2023, smit_rotation_2018}, and \cite{neeleman_kinematics_2021}. However, these studies do not match our observation selection criteria, increasing the uncertainty in the kinematic properties derived when compared to the analysis of REBELS-25.

Overall, our selection results in a total of 37 objects with [CII], [CI], or CO high resolution kinematic studies from the literature. We plot the stellar mass and SFR for the selected galaxies in Figure \ref{fig:comparison sample} at three different redshift ranges ($0.5\leq z <3$, $3\leq z <4.5$, $4.5\leq z < 7$). Due to the limited sensitivities and resolution of observations at high redshift, we see that our selection is biased towards massive ($M_* \gtrsim 10^9 \mathrm{ M_{\odot}}$) main-sequence or starburst (SFR$\gtrsim 10 \mathrm{ M_{\odot}}\mathrm{yr}^{-1}$) galaxies. We note that, as discussed in \cite{hygate_alma_2023}, the definition of a main-sequence and a starburst galaxy differs in the literature. In Figure \ref{fig:comparison sample}, we adopt the \cite{schreiber_herschel_2015} main sequence relation, and define a starburst threshold at 4 times above this line (e.g., \citealt{rodighiero_lesser_2011}). Here, we see that REBELS-25 sits close to the starburst threshold when using our adopted stellar mass from \cite{bouwens_reionization_2022}. However, as discussed in \cite{hygate_alma_2023}, the classification of REBELS-25 can change depending on the SED fitting applied and  stellar mass value adopted.

\subsection{The cosmic evolution of turbulence within galaxies}
\label{sec:evolution of turbulence}

\begin{figure*}
    \centering
    \includegraphics[width=0.9\textwidth]{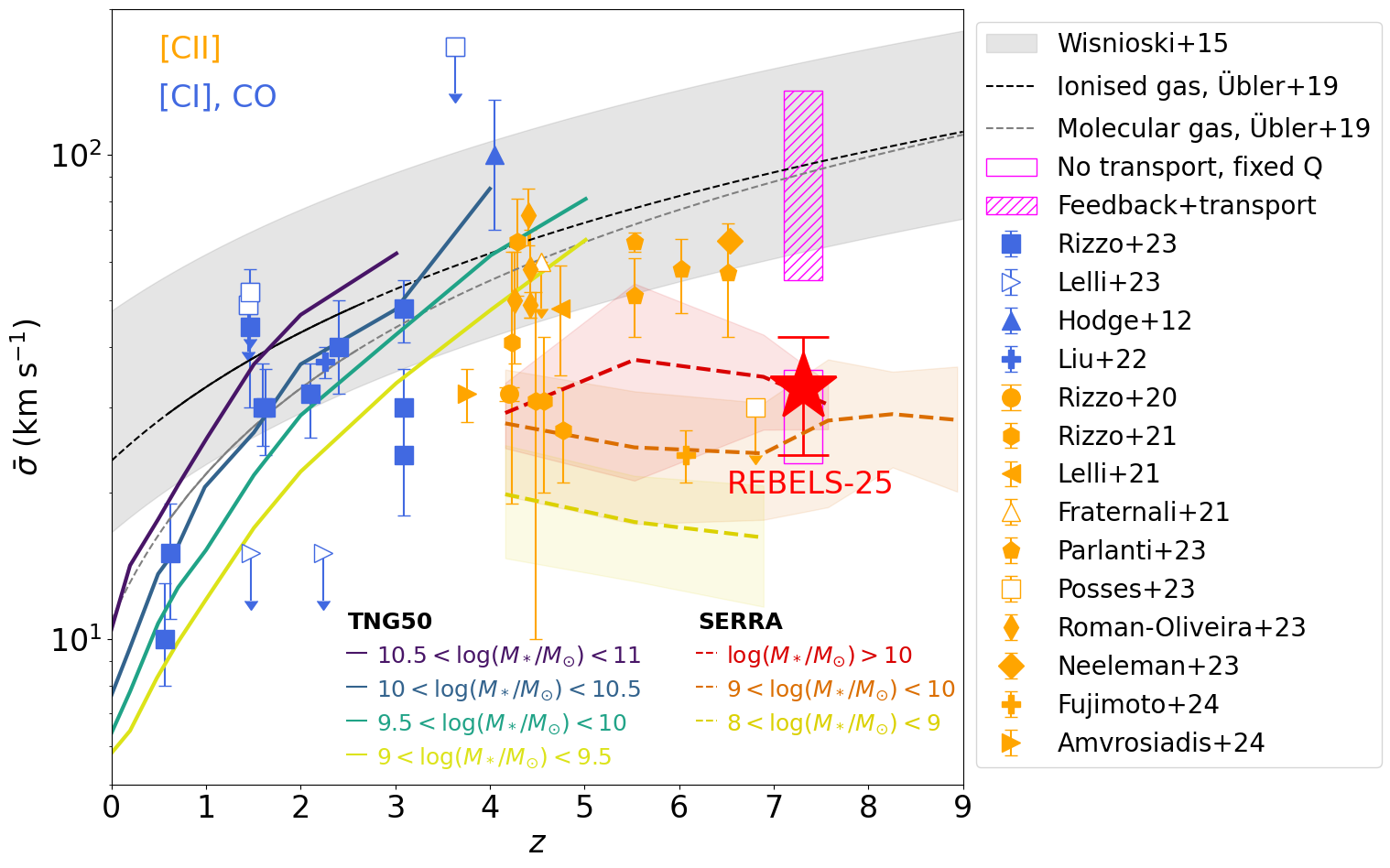}
    \caption {Velocity dispersion as a function of redshift for REBELS-25 (red star-shaped marker) and other well-resolved galaxies from the literature with CO or [CI] (blue) and [CII] (orange) observations, and  The markers are the same as in Figure \ref{fig:comparison sample}. Empty markers show where one or more of the kinematic parameters ($V_{\mathrm{rot}}$ or $\sigma$) is given as an upper or lower limit. The grey shaded region shows the predicted evolution of $\sigma$ from the semi-analytic model of \protect\cite{wisnioski_kmos3d_2015}. The yellow-to-purple solid lines show predictions from Illustris TNG50 (\protect\citealt{pillepich_first_2019}) in increasing mass bins, whilst the yellow-to-red dashed lines and shaded areas show the predictions from the SERRA simulations (\protect\citealt{kohandel_dynamically_2023}) and their $1\sigma$ uncertainties. The empty magenta box shows the prediction from the \protect\cite{krumholz_unified_2018} model that only considers feedback effects as the driver of turbulence, whilst the hatched magenta box shows the prediction from the model that incorporates both feedback effects and gravitational perturbations, as described in Section \ref{sec:evolution of turbulence}.}
    \label{fig:disp vs z}
\end{figure*}

\begin{figure*}
    \centering
    \includegraphics[width=0.9\textwidth]{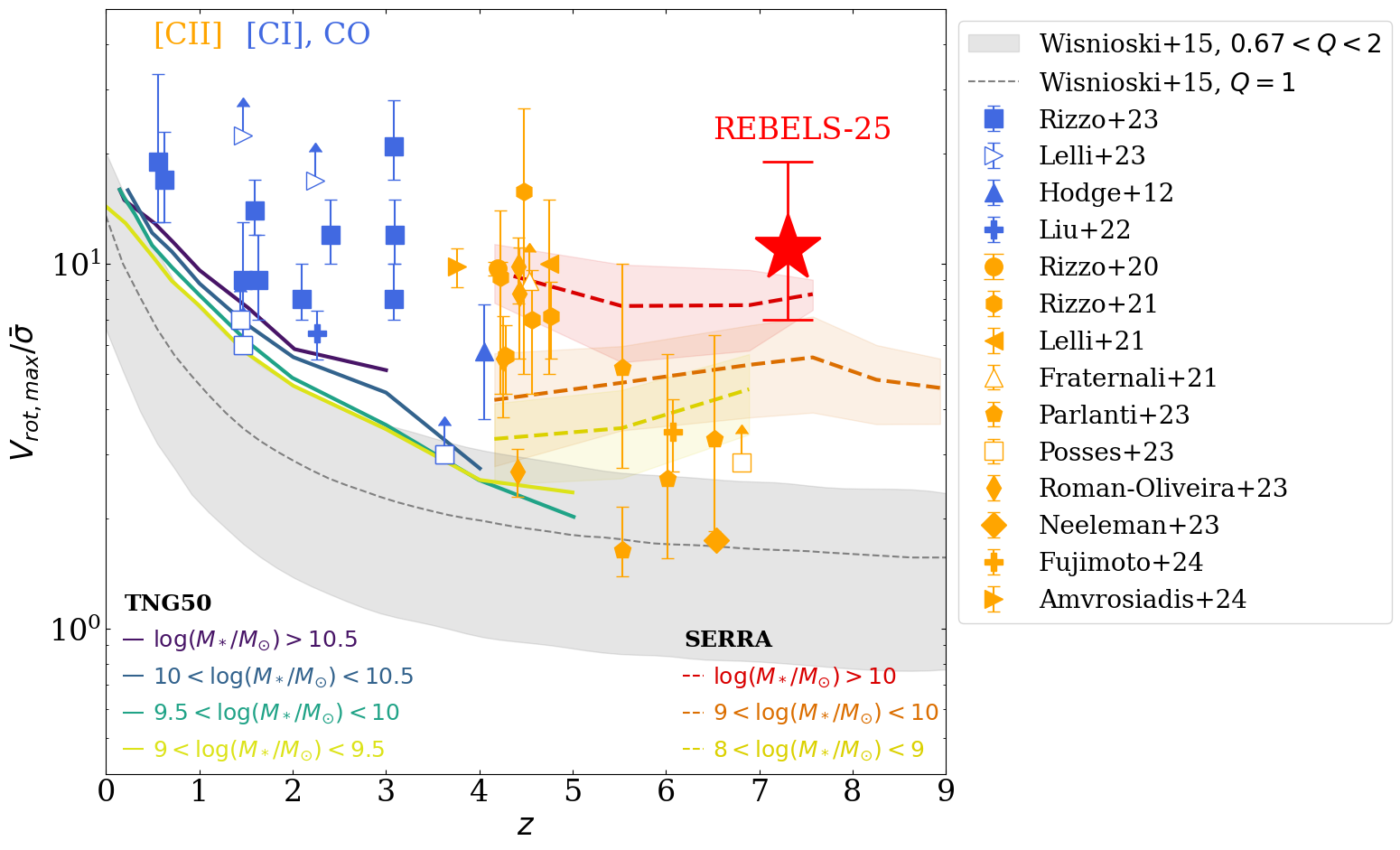}
    \caption {As with Figure \ref{fig:disp vs z}, but instead the ratio of ordered to random motion (the ratio of the maximum rotational velocity to the average velocity dispersion) is plotted as a function of redshift.}
    \label{fig:ratio vs z}
\end{figure*}

In Figure \ref{fig:disp vs z}, we plot the average velocity dispersion, $\bar{\sigma}$, for REBELS-25 and the literature sources as a function of redshift. We plot also the analytical fit to the large IFU survey KMOS3D at $z \sim 1-3$ (\citealt{ubler_evolution_2019}) for both ionised and atomic/molecular gas (solid black and grey lines, respectively) and extrapolate these fits to higher redshifts (dashed black and grey lines). As discussed in Section \ref{sec:intro}, this \cite{ubler_evolution_2019} study of 175 star-forming disc galaxies finds that turbulence increases with redshift for both warm ionised gas and cold neutral gas tracers. \cite{wisnioski_kmos3d_2015} find that this evolution is consistent with being driven by a redshift evolution of the gas fraction according to the Toomre stability criterion (\citealt{toomre_gravitational_1964}). We plot the predictions from this semi-analytical \cite{wisnioski_kmos3d_2015} model for the H$\alpha$ emission of a disc galaxy of mass $\log(M_*/M_{\odot}) \sim 10$ with $Q=1$ by the shaded grey area in Figure \ref{fig:disp vs z}. Additionally, we plot the predictions from IllustrisTNG50 (\citealt{nelson_first_2019}) simulations for H$\alpha$ emission (\citealt{pillepich_first_2019}) and from SERRA simulations (\citealt{pallottini_survey_2022}) for [CII] emission (\citealt{, kohandel_dynamically_2023}) for different mass bins. The predictions from the SERRA simulations are specifically for [CII] emission from mock ALMA observations at different inclinations of 3218 galaxies at $8\leq\log(M_*/M_{\odot})\leq10.3$ from $4<z<9$, with full details of their kinematic analyses given in \cite{kohandel_dynamically_2023}.

We find that the average velocity dispersion of REBELS-25 agrees better with the SERRA simulations than the other predictions at high-$z$. This likely follows from the fact that the \cite{wisnioski_kmos3d_2015} and TNG50 predictions are for warm ionised gas rather than cold gas tracers. \cite{ubler_evolution_2019} find an average difference of 10$-$15 km s$^{-1}$ between velocity dispersions measured from molecular or atomic gas and corresponding measurements from ionised gas at $z\sim1-3$, although this offset may evolve with redshift. Similarly, \cite{kohandel_dynamically_2023} find that $\sigma$ measured from H$\alpha$ emission is $\sim 2\times$ larger than $\sigma$ as measured from [CII] emission. However, following from the expectation that higher redshift galaxies should be dominated by turbulent motions, we would still expect the same general trend with the turbulence increasing with redshift. Instead, we see that the average velocity dispersion for REBELS-25 ($\bar{\sigma} = 33 \pm 9 \mathrm{km s^{-1}}$) is comparable to $z<3$ galaxies.

To investigate why this may be the case, we consider the processes responsible for driving the turbulence, and hence the gas velocity dispersion, within galaxies. Firstly, for REBELS-25, the derived $f_{\mathrm{gas}}=0.93^{+0.03}_{-0.07}$ is consistent with the predictions of \cite{wisnioski_kmos3d_2015} according to its redshift and sSFR, and one may therefore expect it to have a similar velocity dispersion as predicted by the \cite{wisnioski_kmos3d_2015} model (albeit lower for the cold gas tracer used), where the dispersion is given by:

\begin{equation}
\bar{\sigma} = \frac{1}{\sqrt{2}} V_{\mathrm{rot, max}} f_{\mathrm{gas}}(z).
\end{equation}

This would predict a $\bar{\sigma}$ of $\sim 240$ km s$^{-1}$ (for a warm ionised gas tracer, so $\sim 120$ km s$^{-1}$ for the [CII] emission according to \citealt{kohandel_dynamically_2023}). However, as discussed in \cite{wisnioski_kmos3d_2015}, the redshift evolution of the necessary parameters to define $f_{\mathrm{gas}}(z)$, such as the depletion time and the sSFR, are uncertain at $z\gtrsim 3$, and may therefore not be applicable at the high redshift of REBELS-25.

Secondly, we consider the models defined in \cite{krumholz_unified_2018}, which find that the turbulence within galaxies is driven by gravitational instabilities and/or feedback effects. For the galaxies in \cite{parlanti_alma_2023}, from which we have selected seven main-sequence star forming galaxies for comparison with REBELS-25, the models that include both feedback effects and gravitational instabilities from \cite{krumholz_unified_2018} were found to agree well with their data. However, in \cite{rizzo_dynamical_2021}, from which we have plotted the kinematic properties of four starburst galaxies and one main-sequence galaxy, models with only energy injected by feedback effects were found to be sufficient. The same is also true for the four galaxies in \cite{roman-oliveira_dynamical_2024}, which are a mix of main sequence and starburst galaxies. For REBELS-25, we have plotted the predicted velocity dispersion according to both of these \cite{krumholz_unified_2018} models in Figure \ref{fig:disp vs z} (for cool atomic or molecular gas). Here we see that, as with the galaxies in the \cite{rizzo_dynamical_2021} and \cite{roman-oliveira_dynamical_2024} samples, the models considering only feedback effects better agree with the velocity dispersion obtained for REBELS-25. This is perhaps due to the lower mass of the galaxies in the \cite{parlanti_alma_2023} sample in comparison to the \cite{rizzo_dynamical_2021} and \cite{roman-oliveira_dynamical_2024} samples, which could mean that gravitational instabilities have more of an effect on the bulk motion of the galaxies. For REBELS-25, the stellar mass quoted in Table \ref{tab:R25 properties} is comparable to those in the \cite{parlanti_alma_2023} sample, but this value is likely to be an underestimate if there is a large amount of dust obscuration. Improved constraints on the stellar mass will be possible with the upcoming JWST data on this galaxy.

Finally, we also consider that, for this sample of galaxies that is biased towards the most active and massive star-forming galaxies, the difference between the velocity dispersion traced by the warm ionised and cold gas emission may be more extreme than in other studies such as \cite{ubler_evolution_2019}, with potentially more intense outflows and other radial motions that could increase the turbulence in the warm ionised gas. In high-$z$ studies, comparing both cold and warm galaxy kinematics is still limited due to scarce data, although some progress is now being made with the advent of JWST (\citealt{parlanti_ga-nifs_2023, fujimoto_primordial_2024, ubler_ga-nifs_2024}). We note that upcoming analysis of ALMA Cycle 9 Band 8 high resolution observations of the [OIII]88$\mu$m line for REBELS-25 (Algera et al. in prep, ID:2022.1.01324.S) will help to shed light on the effect of different gas tracers on the fitted kinematics of this galaxy.

\subsection{The cosmic evolution of rotational support}

Another kinematic property found to evolve with redshift is the ratio of maximum rotational velocity to average velocity dispersion, $V_{\mathrm{rot, max}}/\bar{\sigma}$. As with Figure \ref{fig:disp vs z}, we plot this ratio as a function of redshift for REBELS-25, the literature sources and various predictions in Figure \ref{fig:ratio vs z}. Here, we see that REBELS-25, and most of the galaxies in our comparison sample, have $V_{\mathrm{rot, max}}/\bar{\sigma} > 2$. This places the majority of these sources above the predictions from the \cite{wisnioski_kmos3d_2015} model and from the TNG50 simulations, although we again note that these predictions are for warm ionised gas tracers, which are likely to result in higher velocity dispersions and therefore lower $V_{\mathrm{rot, max}}/\bar{\sigma}$.

In Figure \ref{fig:ratio vs z}, we again plot the predictions for the evolution of the degree of rotational support from the SERRA simulations (\citealt{kohandel_dynamically_2023}). Here we see that REBELS-25 is consistent within the uncertainties with the predictions for the $\log(M_*/M_{\odot})>9$ sample. This is also the case for most of the galaxies in our comparison sample at $z>4$, although we note that the galaxies from \cite{parlanti_alma_2023} have large uncertainties such that they are consistent with most of the predictions shown here.

The discrepancies between the predictions from SERRA and TNG50 could be due to a combination of reasons. The first possible cause is, as discussed above, the SERRA predictions shown in Figure \ref{fig:ratio vs z} are for kinematics as traced by [CII] emission, whereas the TNG50 predictions are for H$\alpha$ emission. Another possible cause is that SERRA uses mock ALMA observations (\citealt{kohandel_dynamically_2023}) whereas TNG50 takes the intrinsic kinematic properties from a face-on projection of the simulated galaxies (\citealt{pillepich_first_2019}). However, these reasons do not explain why the SERRA predictions show almost no redshift evolution. Instead, the SERRA predictions mainly indicate that the stellar mass of the galaxy has a more significant impact on its degree of rotational support. Similar emphasis on this mass dependence is also made in \cite{dekel_mass_2020}, where both the VELA simulations and SAMI observations indicate a significant increase in the degree of rotational support at $\log(M_*/M_{\odot}) \sim 9$. Similarly, \cite{tiley_kmos_2021} find that stellar mass, rather than redshift, most strongly correlates with the disc fraction amongst star-forming galaxies at $z\sim1.5$, observing only a modest increase in the prevalence of discs between $z \sim1.5$ and $z \sim 0.04$ at fixed stellar mass. Additionally, recent FIRE-2 simulations have implied that disc galaxies can exist at any redshift in sufficiently massive haloes (\citealt{gurvich_rapid_2022}). However, as can be seen from the mass ranges of our comparison sample in Figure \ref{fig:comparison sample}, there is a lack of high resolution observations of cold gas kinematics in low-mass ($M_*<10^9 \mathrm{ M_{\odot}}$) galaxies at high-$z$ to confidently validate these hypotheses. 

\begin{figure}
    \centering
    \includegraphics[width=0.49\textwidth]{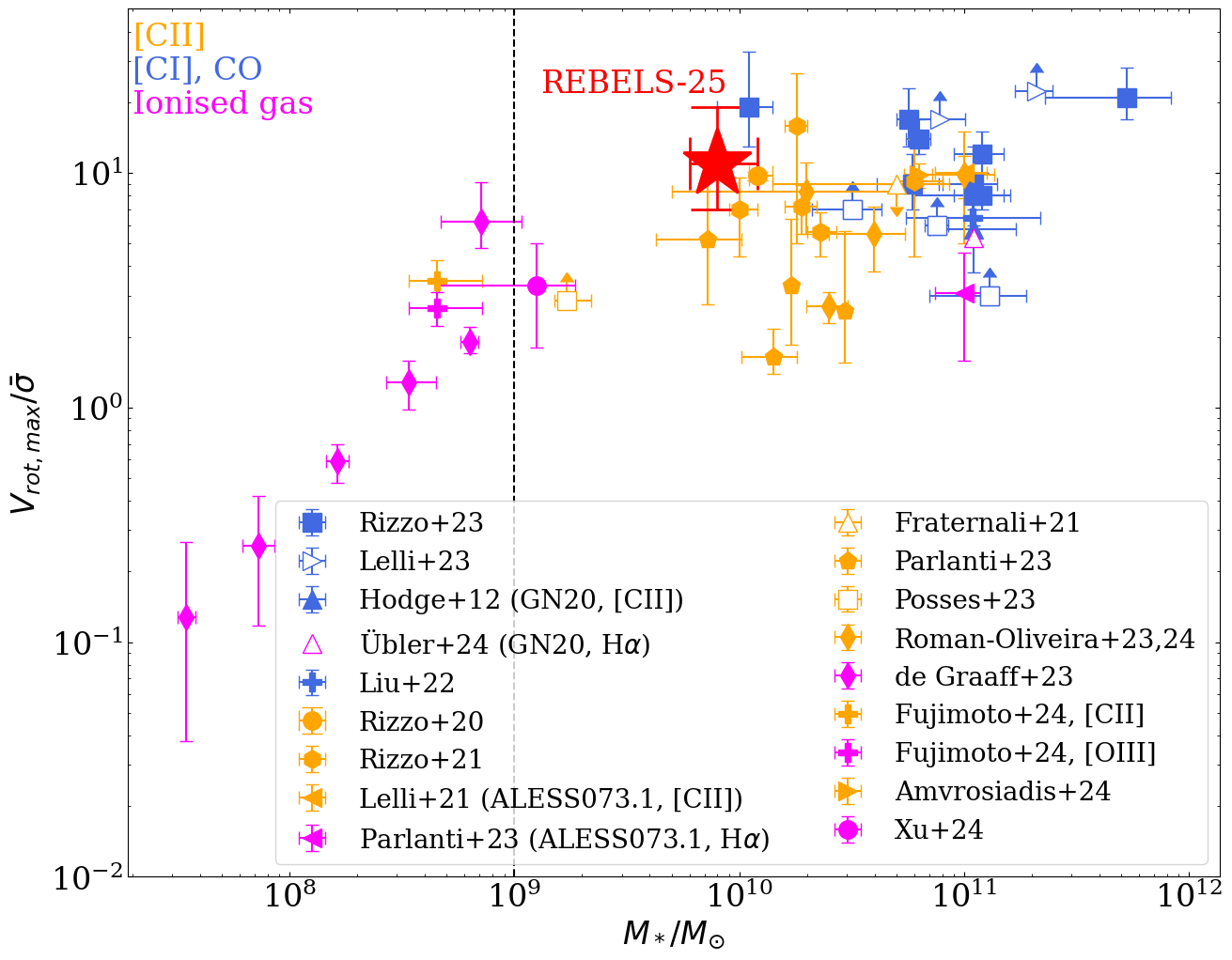}
    \caption {As with Figures \ref{fig:disp vs z} and \ref{fig:ratio vs z}, but instead the ratio of ordered-to-random motion is plotted as a function of stellar mass. The magenta markers are from recent JWST studies of ionised gas kinematics (de \protect\citealt{de_graaff_ionised_2023, fujimoto_primordial_2024,parlanti_ga-nifs_2023,xu_dynamics_2024}). The black vertical dashed line indicates a cut-off stellar mass at $\sim 10^9 \mathrm{ M_{\odot}}$ described in \protect\cite{gurvich_rapid_2022} as a threshold mass where rotationally supported discs can form.}
    \label{fig:ratio vs mass}
\end{figure}

To investigate this mass dependence further, we therefore also make comparisons to a recent study of the ionised gas kinematics at high-$z$ with JWST NIRSpec Multi-Object Spectroscopy (MOS) observations (\citealt{de_graaff_ionised_2023}), and also studies of individual galaxies with NIRSpec IFU observations, including a lensed galaxy at $z=6.072$ (Cosmic Grapes, \citealt{fujimoto_primordial_2024}), ALESS073.1 at $z=4.76$, and GN-z11 at $z=10.60$ (\citealt{xu_dynamics_2024}). We note that due to the compactness of GN-z11, it is not currently possible to rule out a merger scenario from the current spatial resolution (\citealt{xu_dynamics_2024}). Additionally, the observations from \cite{de_graaff_ionised_2023} use the NIRSpec MOS, and so it is also not possible to differentiate between merging and rotating disc systems since the objects are resolved by only a few resolution elements along a single spatial direction. However, the galaxies studied in \cite{de_graaff_ionised_2023} are amongst the lowest mass galaxies studied kinematically at $z>1$ so far, and therefore make for an interesting additional comparison sample. We also add that the kinematics of ALESS073.1 have been studied with both ALMA ([CII], \citealt{lelli_massive_2021}) and with JWST (H$\alpha$, \citealt{parlanti_ga-nifs_2023}) observations at high resolution, as has the 'Cosmic Grapes' $z=6.072$ galaxy from \cite{fujimoto_primordial_2024}. In Figure \ref{fig:ratio vs mass}, we plot the $V_{\mathrm{rot, max}}/\bar{\sigma}$ against stellar mass for the same galaxies as in Figure \ref{fig:ratio vs z}, along with the seven galaxies from \cite{de_graaff_ionised_2023} in magenta diamond markers.

Figure \ref{fig:ratio vs mass} appears to show a tentative positive correlation between $V_{\mathrm{rot, max}}/\bar{\sigma}$ and $M_*$. For the warm ionised gas tracers, it is likely that the velocity dispersions are higher than if they were traced by cold gas (as seen for Cosmic Grapes and ALESS073.1). We could therefore assume that the cold gas velocity dispersion would be lower, which would in turn increase $V_{\mathrm{rot, max}}/\bar{\sigma}$. It is therefore not possible to confidently assess this trend. It should also be noted that there are large uncertainties in deriving the stellar mass at high redshift, which may also affect any observed relationship. Future studies with spatially resolved SED fitting may help to improve stellar mass estimates at high-$z$ (e.g., \citealt{abdurrouf_spatially_2023, gimenez-arteaga_spatially_2023}).

\section{Summary and Conclusions}
\label{sec:conclusions}

In this paper, we have presented follow-up high-resolution ($\sim$ 710 pc) ALMA [CII] and dust ($\sim 150 \mu$m) continuum observations of REBELS-25; a massive star-forming galaxy at $z=7.31$, originally targeted as part of the ALMA REBELS LP (\citealt{bouwens_reionization_2022}). In previous studies (\citealt{stefanon_brightest_2019, hygate_alma_2023}), this galaxy was found to have multiple UV clumps, indicating a potential disturbed, merging or clumpy morphology. However, the follow-up high resolution observations presented here indicate that the [CII] emission in this galaxy is well fit by a near-exponential disc profile, with a Sérsic index of $1.3\pm0.2$. These UV clumps that are offset from the [CII] and dust are therefore likely to be the result of dust obscuration, or are potentially star forming clumps embedded into the gas disc. As well as 2D Sérsic fitting, we have used \texttt{CANNUBI} to fit additional morphological parameters of this galaxy, and find that it has a low inclination ($i= 25\pm6{^\circ}$) and potentially a very thin disc ($Z_0<710$ pc). In addition, we see some evidence of more complex morphological features, including tentative evidence of a bar, identified by fitting elliptical isophotes to both the [CII] and dust continuum emission.

We find the [CII] kinematics of REBELS-25 are well explained by a rotation-dominated disc using the 3D tilted ring fitting tool, \texttt{$^{\mathrm{3D}}$BAROLO}, and several independent criteria to distinguish between a disc and a major merger, including the five-disc criteria from \cite{wisnioski_kmos3d_2015} and \cite{jones_alpine-alma_2021}, and also the PV Split method from \cite{rizzo_dynamical_2021}. The best-fit rotating disc model with \texttt{$^{\mathrm{3D}}$BAROLO} reveals a low overall turbulence ($\bar{\sigma}=33\pm 9$ km s$^{-1}$) and a high ratio of $V_{\mathrm{rot, max}}/\bar{\sigma} = 11^{+8}_{-4}$. This low average dispersion velocity obtained for REBELS-25 is consistent with stellar feedback as the main driver of turbulence within this galaxy, as has also been found for other dynamically cold discs in the high redshift Universe (\citealt{rizzo_dynamical_2021, roman-oliveira_dynamical_2024}). However, we do see some evidence of non-circular motions, which could be due to inflows/outflows, a minor merging component, a central bar and/or spiral arms.

Using the rotational velocity at $r_e$ (equivalently, this is where the curve begins to flatten), we find a total dynamical mass of ($1.2\pm0.3)\times 10^{11} \mathrm{ M_{\odot}}$. This results in a gas mass of $M_{\mathrm{gas, tot}}=(1.1^{+0.4}_{-0.3})\times 10^{11} \mathrm{M_{\odot}}$ and $\alpha_{\mathrm{[CII]}}=62^{+32}_{-22} \mathrm{M_{\odot}/L_{\odot}}$ for $M_* = 8^{+4}_{-2}\times10^9 \mathrm{ M_{\odot}}$, however, the total stellar mass of this galaxy is likely very uncertain thanks to dust obscuration, therefore we find that estimates of $M_{\mathrm{gas}}$ and $\alpha_{\mathrm{[CII]}}$ are currently also highly uncertain. Upcoming JWST observations (GO-1626, GO-6036 and GO-6480) will help improve estimates of the stellar mass and stellar morphology of the galaxy, enabling future work on dynamical modelling and rotation curve decomposition of REBELS-25 with the kinematic information obtained in this paper.

An increasing number of dynamically cold discs have been identified in the high-$z$ Universe, although these observations are often not sufficiently resolved to confidently distinguish between mergers and discs. With these sub-kpc resolution observations, REBELS-25 is amongst the most distant robustly confirmed cold discs observed to date. This finding of a very distant ($z=7.31$), very dynamically cold ($V_{\mathrm{rot, max}}/\bar{\sigma} = 11^{+8}_{-4}$) disc challenges the predictions from many state-of-the-art models and cosmological simulations, which tend to predict very turbulent and dispersion-dominated discs at $z>3$. By comparing to other cold gas kinematics studies of $z>0.5$ galaxies with observational data of similar quality, we find that dynamically cold discs seem to be more common in the high-$z$ Universe than predictions based on warm ionised gas tracers (e.g., \citealt{wisnioski_kmos3d_2015, pillepich_first_2019, ubler_evolution_2019}), although we note an observational bias towards massive star forming galaxies. However, these observations are more consistent with recent predictions from the SERRA cosmological simulations (\citealt{pallottini_survey_2022}) based on mock observations of [CII] emission (\citealt{kohandel_dynamically_2023}), suggesting that the kinematic tracer used significantly impacts the derived velocity dispersion, and therefore degree of rotational support. For the case of REBELS-25, both high resolution ALMA observations of [OIII]88$\mu$m emission (Algera et al. in prep, ID:2022.1.01324.S) and JWST observations of [OIII]5007$\AA$ kinematics (GO-6036 PI J. Hodge, and GO-6480 PI S. Schouws) will enable a comparison between cold and warm ionised gas tracers.

REBELS-25 is a massive ($M_* = 8^{+4}_{-2}\times 10^9 \mathrm{ M_{\odot}}$) galaxy, and several studies have suggested a stronger dependence on the mass of the galaxy, rather than its redshift, in setting the evolution of the gas kinematics (e.g. \citealt{dekel_mass_2020, gurvich_rapid_2022, kohandel_dynamically_2023}). A comparison to lower-mass galaxies tentatively provides observational evidence to support this (Figure \ref{fig:ratio vs mass}), although further work, particularly at lower stellar masses, is necessary to study this mass dependence. JWST will be a powerful tool for such studies, with ionised gas kinematic studies of fainter, lower mass galaxies at high-$z$ now becoming more feasible (\citealt{de_graaff_ionised_2023}).

Overall, in this work we have shown that dynamically settled rotating disc galaxies, such as REBELS-25, can form as early as just 700 Myr after the Big Bang. We therefore expect that future, high resolution studies of cold gas kinematics at high-$z$ will reveal even more cold, massive discs. In particular, ongoing ALMA observations of other REBELS galaxies will enable robust kinematic modelling of additional rotating disc candidates at $z\sim6-8$ (Phillips et al. in prep).

\section*{Acknowledgements}

The authors thank Jorge González-López for their useful discussions about the JvM effect and correction. The authors also acknowledge Mahsa Kohandel for useful discussions relating to the SERRA simulations. JH acknowledges support from the ERC Consolidator Grant 101088676 (VOYAJ). M. R. is supported by the NWO Veni project ”Under the lens” (VI.Veni.202.225). PD acknowledges support from the NWO grant 016.VIDI.189.162 (``ODIN") and from the European Commission's and University of Groningen's CO-FUND Rosalind Franklin program. AF acknowledges support from the ERC Advanced Grant INTERSTELLAR H2020/740120. IDL acknowledges funding support from ERC starting grant 851622 DustOrigin. M.A. acknowledges support from FONDECYT grant 1211951, and ANID BASAL project FB210003. RAAB acknowledges support from an STFC Ernest Rutherford Fellowship [grant number ST/T003596/1]. PEMP acknowledges support from the Dutch Research Council (NWO) through the Veni grant VI.Veni.222.364.

\section*{Data Availability}
The ALMA observations used in this article are available in the ALMA archive https://almascience.eso.org/aq/ and can be accessed with their project codes: 2019.1.01634.L and 2021.1.01603.S. The COSMOS-DASH HST image used available from the Mikulski Archive for Space Telescopes on the COSMOS-DASH page: https://archive.stsci.edu/hlsp/cosmos-dash. The data underlying this article will be shared on reasonable request to the corresponding author.
 



\bibliographystyle{mnras}
\bibliography{REBELS_25_project_v6} 




\appendix

\section{Tests using different [CII] line cubes}

\begin{table*}
\caption[]{Table showing some of the observation details of the different [CII] line cubes tested, and the resulting ratio of ordered to random motion ($V_{\mathrm{rot}}/\bar{\sigma}$).}
\normalsize
\centering
\begin{tabular}{lllllll}
\centering
        
        Measurement set & Channel width & Weighting & Robust & Beam & RMS & $V_{\mathrm{rot, ~max}}/\bar{\sigma}$\\
         &  (km s$^{-1}$) &  &  & ("$\times$") & ($\mu$Jy/beam/channel) & \\
	\hline\hline
    Cycle 8 &  5 & Natural & - & 0.14$\times$0.13& 210 & 10$^{+6}_{-4}$ \\
    Cycle 8 &  10 & Natural & - & 0.14$\times$0.13 & 150 & 9$^{+8}_{-3}$ \\
    Cycle 8 &  15 & Briggs & 0.5 & 0.12$\times$0.10 & 140 & 11$\pm$5  \\
LP $+$ Cy 8, JvM-corr &  20 & Natural & - & 0.15$\times$0.14  & 12$^{*}$  & $8\pm2$  \\

        \hline\hline
	
\end{tabular}
\begin{tablenotes}
\item * Note that the residuals for the JvM-corrected cube are scaled by a factor $\epsilon=0.18$.
\end{tablenotes}
\label{tab:data tests}
\end{table*}

In Table \ref{tab:data tests}, we list the properties of the ALMA [CII] line cubes used to test our main findings. The first three rows correspond to Cycle 8 only line cubes created with a range of channel widths (5, 10 and 15 km s$^{-1}$) and using both natural and Briggs-weighted data sets. For the measurement set labelled "LP + Cy 8, JvM-corr", we concatenate the low resolution REBELS LP data with the high resolution Cycle 8 observations. We follow the same reduction steps as outlined in Section \ref{sec:reduction} for each data set, however for the combined cube we add additional steps, which are described below, to account for the different array configurations used in the two observations. 

For the combined cube, we are impacted by the JvM effect, as mentioned in Section \ref{sec:reduction}. This effect occurs when the volume of the \texttt{CLEAN} beam differs from the volume of the dirty beam, resulting in an incorrect flux scaling of the \texttt{CLEAN} residual map. Consequently, if no correction is applied, we derive a [CII] flux from the concatenated dataset that is $\sim 3.4$ times higher than the [CII] flux from the LP data (\citealt{hygate_alma_2023}) alone. Recently, a correction for this effect has been suggested by \cite{czekala_molecules_2021}. In summary, a correction factor, $\epsilon$, is determined from the ratio of the dirty to \texttt{CLEAN} beam volume (taken at the first null of the dirty beam), and the residuals are then scaled by this factor before being added to the convolved \texttt{CLEAN} model to produce the final imaging.

As in \cite{posses_alma-cristal_2024}, we find that cleaning down to a threshold of $1\sigma_{\mathrm{RMS}}$ minimises this effect, and we find that binning to a higher channel width (20 km s$^{-1}$) also reduces this effect. Since these observations were taken over two years apart and with multiple array configurations, we also apply the \texttt{statwt} task to the continuum-subtracted measurement set to re-assign weights based on the measured noise. For this \texttt{statwt} task, we again fit only line-free channels, since emission could contaminate the weight calculation. We then apply the JvM correction with $\epsilon= 0.18$, as determined from the ratio of the \texttt{CLEAN} and dirty beam volumes. Applying this correction results in a consistent total [CII] flux (in a circular aperture with a radius of 1.75")  between the combined cube, the Cycle 8 data only cubes, and the LP data. However, as discussed in, for example, \cite{casassus_variable_2022, posses_alma-cristal_2024}, the application of the JvM correction means that the residuals have been rescaled, and so the RMS is no longer a true representation of the sensitivity of the observations. This can therefore significantly exaggerate the SNR (\citealt{casassus_variable_2022}). When fitting the morphology and kinematics on the concatenated cube with \texttt{CANNUBI} and \texttt{$^{\mathrm{3D}}$BAROLO}, we therefore increase the \texttt{GROWTHCUT} parameter to 6. We note that the uncertainties on the fitted parameters for the concatenated cube are likely underestimated.

\begin{figure}
    \centering
    \includegraphics[width=0.45\textwidth]{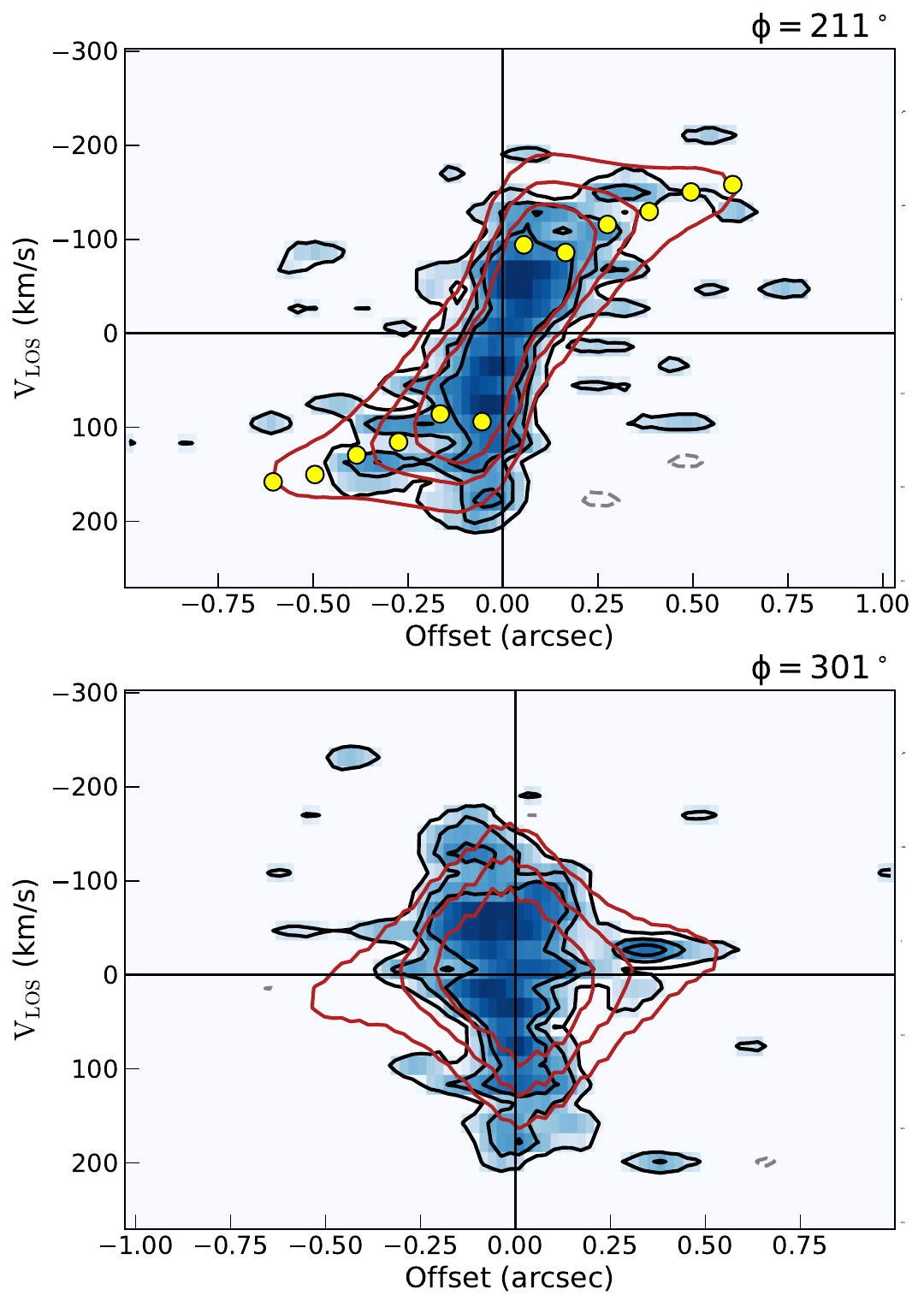}
    \caption {The major-axis (top) and minor-axis (bottom) PVDs for the best-fit \texttt{$^{\mathrm{3D}}$BAROLO} model to the JvM corrected cube. The contours for the data (in black) and the model (in red) are at 6, 12 and 18 $\sigma_{\mathrm{RMS}}$ as calculated from \texttt{$^{\mathrm{3D}}$BAROLO}. Negative contours for the data at  -12 and -6 $\sigma_{\mathrm{RMS}}$ are indicated by the dashed grey contours. Note, however, that this $\sigma_{\mathrm{RMS}}$ is not a true representation of the sensitivity of the data due to the scaling of the residuals in the JvM correction. The yellow markers in the top panel indicate the separation of the six rings fitted by \texttt{$^{\mathrm{3D}}$BAROLO} and their fitted rotation velocities along the line-of-sight. The asymmetry in the model of the minor-axis PVD is the result of a changing PA across the disc which is preferred for this fit.}
    \label{fig:jvm pvds}
\end{figure}

We run some initial morphology fitting tests for each data set with \texttt{Sersic2D} and \texttt{CANNUBI}. Notably, we find a Sérsic index of $\sim 1$ for all cubes tested, and we find the bright, misaligned component discussed in Section \ref{sec:potential features} is present in all of the different cubes tested.  We overall find morphological parameters that are consistent with those in Section \ref{sec: cii morphology} and Table \ref{tab:R25 parameters}, and we therefore fix all the morphological parameters to those in Table \ref{tab:R25 parameters} when fitting with \texttt{$^{\mathrm{3D}}$BAROLO}. For these kinematic fits, we leave only $V_{\mathrm{rot}}$, $\sigma$, and PA$_{\mathrm{kin}}$ as free parameters, as done with the final kinematic fitting in Section \ref{sec:kinematic modelling set up}. For the cube binned to 5 km s$^{-1}$ and for the cube with Briggs weighting, we fit only 4 rings due to the lower SNR. The $V_{\mathrm{rot, ~max}}/\bar{\sigma}$ and their uncertainties are then derived using the same method as in Section \ref{sec:kinematic analysis}, and are listed in Table \ref{tab:data tests}, where we see that all results are consistent with the main findings of this paper. We also show in Figure \ref{fig:jvm pvds} the PVDs for the kinematic fit to the JvM-corrected cube. In all rotation curves produced from these tests, the bump feature, where there is a higher $V_{\mathrm{rot}}$ value in the first ring, is present. From these tests, we therefore conclude that the kinematic parameters we have derived in the main body of the paper are robust. 

\section{\texttt{CANNUBI} corner plots}

\begin{figure*}
    \centering
    \includegraphics[width=\textwidth]{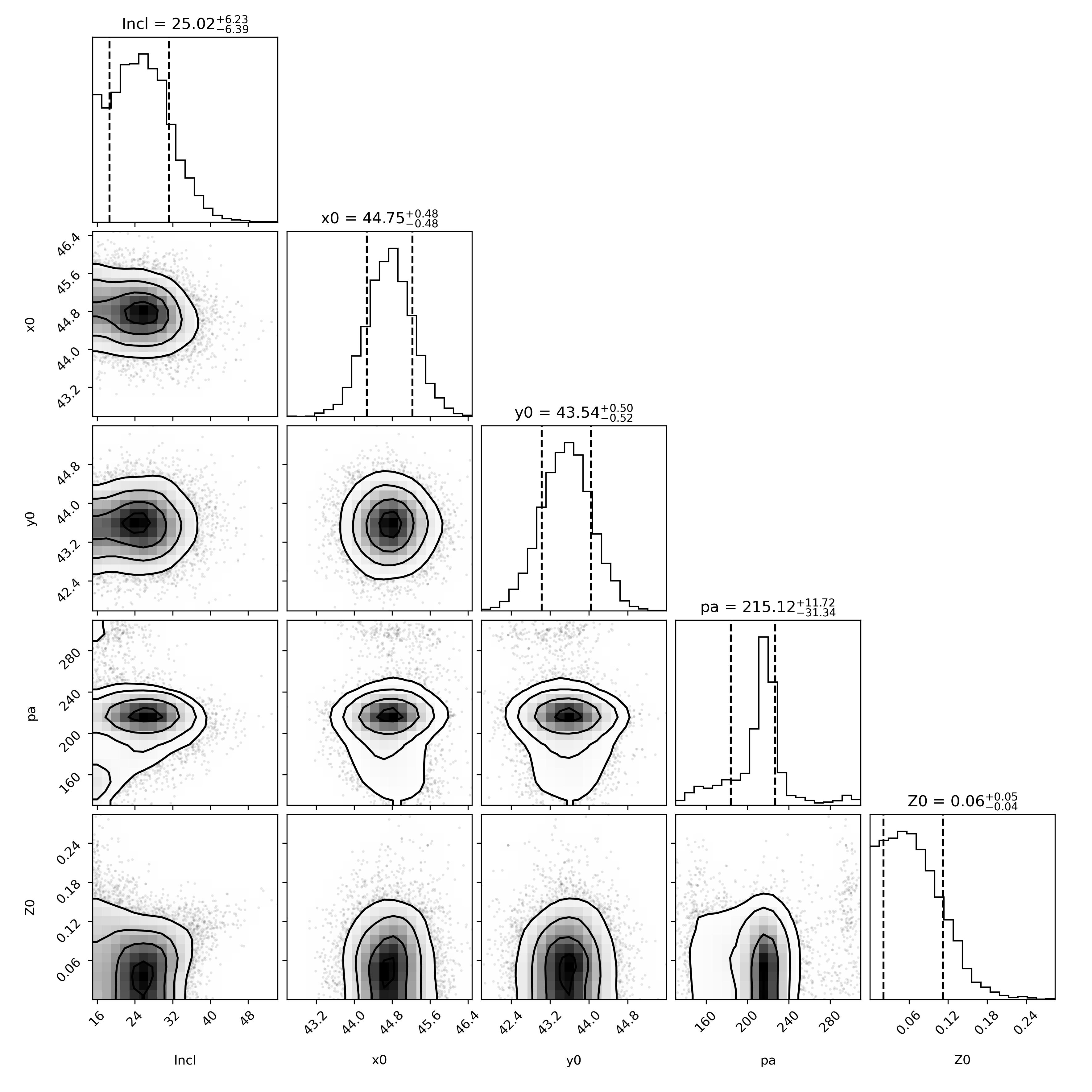}
    \caption {A triangle plot of the posterior distributions for the morphological parameters fitted by \texttt{CANNUBI}.}
    \label{fig:cannubi}
\end{figure*}

In Figure \ref{fig:cannubi}, we show the posterior distributions obtained by applying \texttt{CANNUBI} to our [CII] observations of REBELS-25. The inclination (incl) and position angle (pa) are given in degrees, the disc thickness (Z0) in arcseconds and the geometric centre (x0 and y0) in arbitrary pixel units that are equivalent to J2000 RA, Dec of 10:00:32.3397, +1:44:31.167.

\section{Channel maps}

\begin{figure*}
    \centering
    \includegraphics[width=0.8\textwidth]{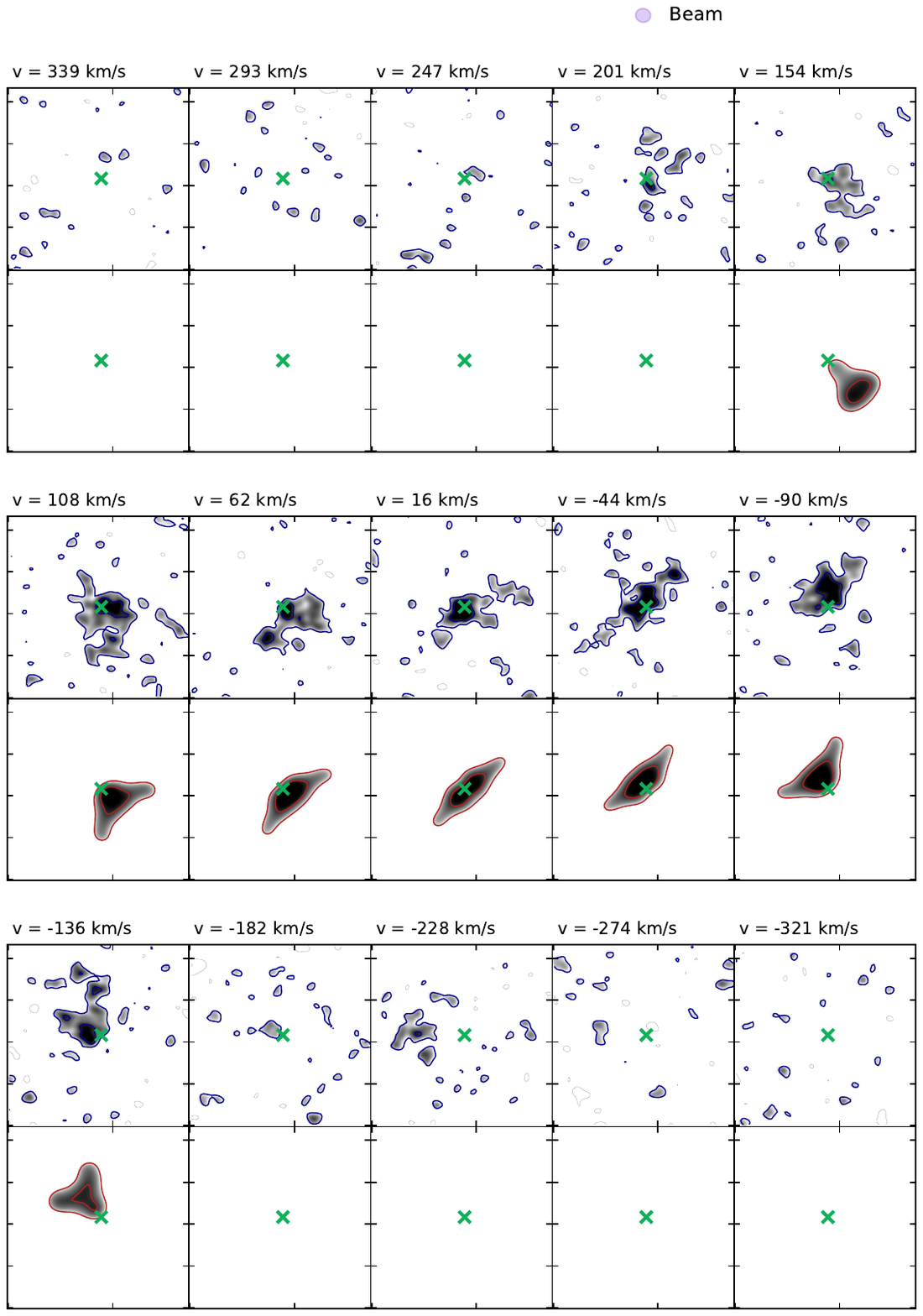}
    \caption {Representative channel maps of REBELS-25, with contours at 2 and 4 $\sigma_{\mathrm{RMS}}$ for the data in blue (upper panels) and the model in red (lower panels). Negative contours at -2$\sigma_{\mathrm{RMS}}$ are indicated by the grey dashed contours. As with Figure \ref{fig:PVDs}, $\sigma_{\mathrm{RMS}}=168\mu$Jy beam$^{-1}$ and is equal to one standard deviation above the median value as calculated by \texttt{$^{\mathrm{3D}}$BAROLO}. The centre of the galaxy as determined from \texttt{CANNUBI} and fixed in the kinematic fit is marked by a green cross.}
    \label{fig:channel maps}
\end{figure*}

In Figure \ref{fig:channel maps}, we show representative channel maps of the [CII] emission of REBELS-25 (top panels) and of the best-fit kinematic model (bottom panels) obtained with \texttt{$^{\mathrm{3D}}$BAROLO}. We show the data in the upper panels with blue contours for the positive emission and grey, dashed contours for the negative emission. The model is shown with red contours in the lower panels. The contours follow the emission intensity according to 2$\sigma_{\mathrm{RMS}}$ and 4$\sigma_{\mathrm{RMS}}$, with the negative contours at $-2\sigma_{\mathrm{RMS}}$. The green cross shows the kinematic centre fitted by \texttt{$^{\mathrm{3D}}$BAROLO}.

\bsp	
\label{lastpage}
\end{document}